\documentclass[preprint,11pt,preprintnumbers,nofootinbib]{revtex4}
\usepackage{graphicx}% Include figure files
\usepackage{dcolumn}% Align table columns on decimal point
\usepackage{bm}% bold math
\usepackage{amsfonts}
\begin{document}

\date{\today}
\title{A Terminal Velocity on the Landscape: \\ Particle Production  near Extra Species Loci in Higher Dimensions}
\author{Diana Battefeld} 
\email[email: ]{dbattefe(AT)princeton.edu}
\author{Thorsten Battefeld}
\email[email: ]{tbattefe(AT)princeton.edu}
\affiliation{Princeton University, Department of Physics, NJ 08544, USA}

\pacs{}
\begin{abstract}
We investigate particle production near extra species loci (ESL) in a higher dimensional field space and derive a  speed limit in moduli space at weak coupling.  This terminal velocity is set by the characteristic ESL-separation and the coupling of the extra degrees of freedom to the moduli, but it is independent of the moduli's potential if the dimensionality of the field space is considerably larger than the dimensionality of the loci, $D\gg d$. Once the terminal velocity is approached, particles are produced at a plethora of nearby ESLs, preventing a further increase in speed via their backreaction. It is possible to drive inflation at the terminal velocity, providing a generalization of trapped inflation with attractive features: we find that more than sixty e-folds of inflation for sub-Planckian excursions in field space are possible if ESLs are ubiquitous, without fine tuning of initial conditions and less tuned potentials. We construct a simple, observationally viable model with a slightly red scalar power-spectrum and suppressed gravitational waves; we comment on the presence of additional observational signatures originating from IR-cascading and individual massive particles. We also show that moduli-trapping at an ESL is suppressed for $D\gg d$, hindering dynamical selection of high-symmetry vacua on the landscape based on this mechanism.

\end{abstract}
\maketitle
\newpage

\tableofcontents

%%%%%%%%%%%%%%%%%%%%%%%%%%%%%%%%%%%%
%%%%%%%%%%%%%%%%%%%%%%%%%%%%%%%%%%%%
\section{Introduction}

The presence of many light fields, or moduli, is a common feature of string theory. At late times, the expectation values of these fields determine low energy observables; hence, their evolution is heavily constraint from the time of nucleosynthesis, although they are expected to be dynamical in the very early universe. For these reasons, moduli trapping is an important aspect of inflationary model building in string theory, see \cite{HenryTye:2006uv,Cline:2006hu,Burgess:2007pz,McAllister:2007bg,Baumann:2009ni} for reviews; often all but one degree of freedom are stabilized by construction during inflation, but how they stabilize is seldom addressed. 

A possible dynamical stabilization process is the String Higgs effect \cite{Bagger:1997dv,Watson:2004aq}, which singles out certain locations in moduli space. At these loci additional degrees of freedom become light, often due to the presence of an enhanced symmetry, and are produced \cite{Kofman:2004yc,Watson:2004aq} if the moduli approach the location. The process of particle production is identical to the one examined in preheating \cite{Traschen:1990sw,Kofman:1997yn}, see \cite{Bassett:2005xm,Kofman:2008zz} for reviews. Backreaction of these new states on the moduli can be dramatic: moduli can be trapped \cite{Kofman:2004yc,Watson:2004aq,Patil:2004zp,Patil:2005fi,Cremonini:2006sx}, or, if moduli drive inflation, the inflaton velocity can decrease temporarily \cite{Chung:1999ve}, as in trapped inflation \cite{Kofman:2004yc,Green:2009ds,Silverstein:2008sg}. A concrete realization of the string Higgs effect can be found in a system of parallel $D$-branes \cite{Witten:1995im}, whose separation is a modulus field from the four-dimensional point of view. If they come close to each other, strings stretching between the branes become massless, gauge symmetries are enhanced and the branes stick to each other due to backreaction -- the modulus is trapped. For a small selection of other examples, see \cite{Seiberg:1994rs,Seiberg:1994aj,Intriligator:1995au,Strominger:1995cz,Witten:1995ex,Katz:1996ht,Bershadsky:1996nh,Witten:1995gx}.

In this paper, we examine the consequences of particle production near such extra species loci (ESL) if the dimensionality of moduli space $D$ is large compared to the dimensionality of the locus, $d$. $D\gg 1$ is natural in string theory, leading to the notion of a landscape \cite{Susskind:2003kw}, yet the effect of ESLs has primarily been investigated for $D-d=1,2$ \cite{Kofman:2004yc,Watson:2004aq,Greene:2007sa}. Based on geometric arguments, we show in Sec.~\ref{sec:particleproduction} that trapping is suppressed if the dimensionality of moduli space is larger than the dimensionality of a given locus ($D>d+1$). This result is expected, since there is no classical attraction towards ESLs and it is improbable to run head on into an ESL if $D\gg d$. This means  that the mere presence of ESLs does not guarantee a dynamical preference of high symmetry states for moduli \cite{Kofman:2004yc,Dine:2000ds,Dine:1998qr}. However, in Sec.~\ref{sec:particpleproductionwithimpactparameter} we show that the presence of many loci with a characteristic inter-ESL distance $x$ leads to a general speed limit, or terminal velocity, on moduli-space. At strong coupling, speed limits are known \cite{Silverstein:2003hf}, leading to DBI-inflation \cite{Alishahiha:2004eh}. Here, we derive a speed limit at weak coupling caused by the combined backreaction of particles produced at many ESLs in the vicinity of the trajectory. 

We take a bottom up approach, assuming the viability of low-energy effective field theory and treating the characteristic distance between ESLs, as well as their dimensionality, as free parameters. Furthermore, we model the additional light degrees of freedom by a massless scalar field that couples to the moduli via interactions of the type $g^2\chi^2\varphi^2$, as in \cite{Kofman:2004yc}. In this notation, the speed limit takes the simple form $|\dot{\vec{\varphi}}|<gx^2\equiv v_t$, given that the classical trajectory is reasonably straight (we allow for a classical potential for the moduli)and $D$ is large. The presence of a terminal velocity offers the opportunity to drive inflation by moduli with hitherto unsuitable potentials. 

This type of inflation is a generalization of trapped inflation \cite{Kofman:2004yc,Green:2009ds}, which recently re-surfaced as monodromy inflation \cite{Silverstein:2008sg}, but with several crucial differences: firstly, particle production at any given ESL is minor because ESLs are not approached too closely, in contrast to the one-dimensional case in \cite{Green:2009ds}; there, extra species points (ESPs) are encountered head on and all particle production occurs at a single ESP at any given time. Secondly, in the large $D$ limit the velocity at which the trajectory is traversed is independent of the slope of the potential as long as the slope is steep enough. Again, this differs from \cite{Green:2009ds}, where the slope still determines the speed and observational parameters, such as the scalar spectral index.

We investigate this type of trapped inflation in higher dimensions  in Sec.~\ref{sec:trappedinfllargeD}, where we encounter several attractive features: focusing on a model with quadratic potentials for the moduli, more than sixty e-folds of inflation can be achieved with sub-Planckian field excursions if $v_t$ is small enough. The inflationary scale is set by the COBE normalization, but the dynamics is not sensitive to the actual shape of the potential as long as a single, reasonably straight classical trajectory with large slope is present. This is, in a sense, opposite to setups needed for slow-roll inflation, where the potential needs to be shallow over long ranges in field space. As a consequence, the $\eta$-problem is alleviated. Furthermore, no fine-tuning of the initial speed is needed: if the initial speed is large, particle production is strong, causing strong backreaction and temporary trapping; after the produced particles are diluted away due to the expansion of the universe, the speed picks up again and the terminal velocity is approached from below. In this sense, trapped inflation is an attractor solution. 

The ongoing particle production during inflation has several additional observational consequences, Sec.~\ref{sec:IR-cascading} and \ref{sec:massive}. Since $\chi$-particles scatter off the inflaton condensate, they cause IR-cascading and additional contributions to the power-spectrum \cite{Barnaby:2009mc,Barnaby:2009dd}; furthermore, $\chi$-particles dilute and become massive once the trajectory moves away from their ESL of production. Thus, one needs to consider the effect of individual massive particles that are coupled to the moduli, which could lead to additional circular cold-spots in the CMBR \cite{Itzhaki:2008ih,Fialkov:2009xm,Kovetz:2010kv}. We plan to investigate both effects, as well as the generation of non-Gaussianities, in a future publication, accompanied by a more rigorous string theoretical implementation of trapped inflation in higher dimensions.

Readers familiar with the notion of particle production, as discussed in \cite{Kofman:2004yc}, and a primary interest in the terminal velocity and consequences for inflation may skip Sec.~\ref{sec:particleproduction} and go directly to the main part of this work, Sec.~\ref{sec:moduliareinflatons}. On the other hand, readers with a primary interest on a dynamical selection principle on the landscape will find Sec.~\ref{sec:particleproduction} and Appendix \ref{A:trappingrw} useful (these sections review \cite{Kofman:2004yc} and contain minor extensions and consequence which are implied to in \cite{Kofman:2004yc} but not fully spelled out).

\section{Quantum Trapping in Moduli Space \label{sec:particleproduction}}
When are moduli affected by the presence of extra species loci? After estimating the distance that moduli can move classically in Sec.~\ref{sec:length}, we review the notion of quantum moduli trapping in Sec.~\ref{sec:trapping} as developed in \cite{Kofman:2004yc} and estimate the trapping probability in Sec.~\ref{sec:trappingprobability}. We comment on drifting moduli in Sec.~\ref{sec:driftingmoduli} and Appendix \ref{A:trappingrw}. Readers familiar with \cite{Kofman:2004yc} may want to skip Sec.~\ref{sec:length} and Sec.~\ref{sec:trapping}.

\subsection{Length of the Field-Trajectory\label{sec:length}}
Let's start by reviewing the effect of Hubble damping on the evolution of moduli fields, as discussed in \cite{Kofman:2004yc}. Consider a $D$-dimensional moduli space with freely moving fields ($\varphi_i$, $i=1\dots D$),  
that is, without a potential that could trap the fields classically. Additionally, assume a flat moduli space metric, leading to a straight trajectory. Defining $\varphi_{cl}$ as the effective field along this trajectory, its equation of motion is
\begin{eqnarray}
\ddot{\varphi}_{cl} +3H\dot{\varphi}_{cl}=0\,, \label{eomphicl0}
\end{eqnarray} 
where a dot denotes a derivative with respect to cosmic time and $H=\dot{a}/a$ is the Hubble parameter. For a constant equation of state parameter $w=\rho/p$ the Hubble parameter becomes $H=\beta/t$ with $\beta\equiv 2/(3+3w)$, and (\ref{eomphicl0}) can be integrated to
\begin{eqnarray}
\dot{\varphi}_{cl}=v\left(\frac{t_0}{t}\right)^{3\beta}\,, \label{eomphicl}
\end{eqnarray}
where $v\equiv \dot{\varphi}_{cl}(t_0)>0$. 

If the universe is dominated by freely moving scalar fields, that is if $w=1$, the length of the field trajectory 
\begin{eqnarray}
s\equiv \left|\varphi_{cl}(t)-\varphi_{cl}(t_0)\right|
\end{eqnarray}
increases logarithmically
\begin{eqnarray}
s = vt_0 \ln\left(\frac{t_0}{t}\right)\,. \label{sincreaseln}
\end{eqnarray}
On the other hand, if $w<1$ that is $\beta>1/3$, equation (\ref{eomphicl}) yields
\begin{eqnarray}
s= -\frac{vt}{3\beta-1}\left(\frac{t_0}{t}\right)^{3\beta}+\frac{vt_0}{3\beta-1}\,, \label{sincreasebeta}
\end{eqnarray} 
which is bounded from above  by $s\lesssim vt_0/(3\beta-1)$. Since $w<1$, we have $v^2/2<3M_p^2H^2$ so that $s<\sqrt{6}\beta M_p/(3\beta-1)$. 

To get intuition for the expected values of $s$ in realistic scenarios, let's consider concrete numbers; in an inflating universe with $\beta \rightarrow \infty$ the upper bound becomes $s<\sqrt{6}M_{p}/3$, less than $M_{p}$ regardless of the initial velocity. However, if $\beta$ is closer to $1/3$, the length of the trajectory can increase above $M_{p}$. For instance, consider a phase dominated by kinetic energy,
\begin{eqnarray}
 \frac{v(t_0)^2}{2\rho(t_0)}\equiv 1-\varepsilon\,,
 \end{eqnarray}
  with $\varepsilon\ll 1$, where a small contribution to the total energy density $\rho$ redshifting as $\rho_c(t)\equiv\varepsilon \rho(t_0)(t_0/t)^{1+w_c}$ (slower than $v^2\propto a^{-6}\propto t^{-2}$) is also present. $\rho_c$ takes over at $t_c=t_0\varepsilon^{-1/(1-w_c)}$ when $v^2(t_c)/2=\rho(t_c)/2$. Using (\ref{sincreaseln}) and (\ref{sincreasebeta}), we get a total path length in the limit $t \rightarrow \infty$ of
\begin{eqnarray}
s= -\frac{1}{1-w_c}vt_0 \ln(\varepsilon)+\sqrt{3}M_{pl}\frac{2}{3(1-w_c)}\,.
\end{eqnarray}   
Thus, an extended kinetic phase is needed to achieve super-Planckian values for $s$. For example, if moduli have an initial Planckian kinetic energy ($v\sim \sqrt{2}M_p^2$) and if they dominate until high-scale slow-roll inflation sets in around $H_{inf}\sim 10^{-5} M_p$ ($w_c \approx -1$), we get $\varepsilon \sim 3 \times 10^{-10}$ and
$s\sim -\sqrt{1/6}M_p \ln(\varepsilon)\sim 9 M_p$. Conceivably,  moduli could also dominate after inflation ended i.e.~down to the SUSY breaking scale $\rho_e\sim TeV^4$, so that $\varepsilon \sim 3\times 10^{61}$ and $s\sim \mathcal{O}(10^2) M_p$.  

In summation,  if the kinetic energy of the moduli does not dominate the energy density of the universe,  the moduli  travel a  limited stretch in field space until they come to rest due to Hubble friction, $s\sim M_p$; for instance, they do not travel far during inflation or in a radiation/matter dominated universe \footnote{They could also be displaced  by quantum mechanical fluctuations during inflation, see Appendix \ref{sec:shapeofarandomwalk}.}. If the moduli's kinetic energy dominates, i.e. before a standard high scale inflationary phase or in the interval after inflation but before SUSY breaking, they can travel  farther ($s_{\mbox{\tiny {before inf.}}}\lesssim 10 M_p$ and $s_{\mbox{\tiny{after inf.}}}\lesssim \mathcal{O}(10^2) M_p$). In the following sections we keep  in mind the interval 
\begin{eqnarray}
1\lesssim s/M_p\lesssim 10^2  \,.
\end{eqnarray}

\subsection{Quantum Trapping \label{sec:trapping}}
Even in the absence of a classical potential, moduli can be trapped near locations where additional particle species become light. These additional degrees of freedom are produced quantum mechanically if the field trajectory gets close enough to an extra species' locus (ESL), and their backreaction can force the trajectory in an orbit around it. Since Hubble damping in an expanding universe causes an inward spiraling trajectory, the moduli get trapped. The origin of the additional light degrees of freedom are often enhanced symmetries at the locus \cite{Kofman:2004yc,Watson:2004aq}, so that ESL or ESP can also denote enhanced symmetry locus or point. 
Quantum moduli trapping was proposed in \cite{Kofman:2004yc,Watson:2004aq} and was recently investigated in \cite{Greene:2007sa}. A field theoretical description of this phenomenon is identical to the resonant particle production after inflation, usually referred to as preheating \cite{Traschen:1990sw,Kofman:1997yn}, see i.e.~\cite{Bassett:2005xm,Kofman:2008zz} for reviews. Our first goal is to estimate a characteristic distance to the ESL below which moduli trapping is common. 

\begin{figure*}[tb]
\includegraphics[scale=0.4,angle=0]{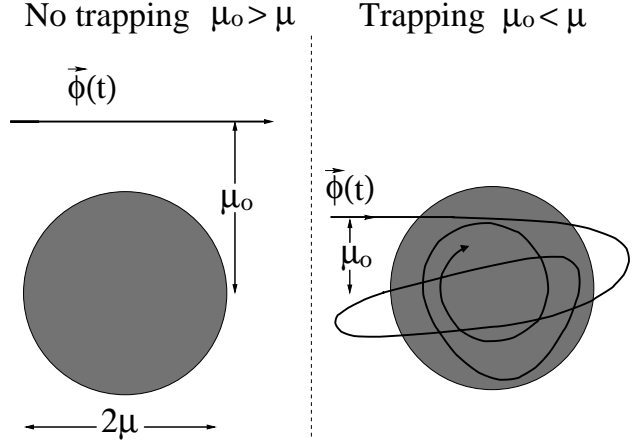}
   \caption{\label{pic:trapping} Schematic of a trapping event: if the trajectory $\vec{\varphi}(t)$ comes close to an ESL (gray circle), that is if the impact parameter $\mu_0$ is smaller than the ESLs effective radius $\mu\lesssim \sqrt{v/g}$, light particles are produced; their backreaction on $\varphi_i$,  along with Hubble damping, can trap the trajectory in the vicinity of the ESL.}
\end{figure*}

Consider a single ESP at the origin where additional light states appear. We model this situation by introducing an new scalar field $\chi$ coupled to the freely moving moduli via a quadratic interaction
\begin{eqnarray}
\mathcal{L}=\frac{1}{2}\sum_{i=1}^{D}\partial_\mu \varphi_i \partial^\mu \varphi_i + \frac{1}{2} \partial_\mu \chi \partial^\mu \chi - \frac{g^2}{2}\chi^2 \sum_{i=1}^{D}\varphi_i^2\,,
\end{eqnarray}
without any bare $\chi$ mass, as in \cite{Kofman:2004yc}. Classically, $\chi=0$ along with a straight trajectory in moduli space is a solution. Starting with this solution,  the trajectory and the ESP span a plane in moduli space.  Hence, we can redefine the fields so that the straight trajectory takes the simple form 
\begin{eqnarray}
\varphi_1&=&v t\\
\varphi_2&=&\mu_0\\
\varphi_i&=&0\,\,\ \mbox{for}\,\,\, i\geq 3 
\end{eqnarray}
where we ignored Hubble friction  so that $v=const$. Hence, to discuss the effect of a single ESL, the problem reduces to the two dimensional case investigated in \cite{Kofman:2004yc}. Efficient particle production in a Fourier mode $\chi_k$ occurs if the non-adiabaticity parameter
\begin{eqnarray}
\omega(t)\equiv \left( k^2+g^2\sum_{i=1}^{D}\varphi^2_i(t)\right)^{1/2}\label{adiabaticity}
\end{eqnarray}
satisfies \cite{Kofman:1997yn}
\begin{eqnarray}
\frac{\dot{\omega}}{\omega^2}>1\,.\label{adiabaticity1}
\end{eqnarray}
Since the effective mass of the $\chi$-field is given by $m_\chi^2(t)=g^2\sum_i\varphi_i^2(t)$ we see that this requirement is satisfied if the $\chi$-field is light. Once $\chi$-particles are produced, they induce a classical confining potential for the moduli, which gets stronger with successive bursts of particle production. Evaluating (\ref{adiabaticity1}) at the point where the ESP is closest, we conclude that trajectories with impact parameters smaller than 
\begin{eqnarray}
\mu_0\lesssim \sqrt{\frac{v}{g}}\equiv \mu \label{defmu}
\end{eqnarray}
have a strong chance to get trapped near the ESP (see Fig.~\ref{pic:trapping} for a schematic). The value of $g$, and correspondingly  the value of the characteristic impact parameter $\mu$ below which trapping is common, is model dependent. In the case of a higher dimensional ESL the discussion is analogous, with $\mu_0$ denoting the minimal distance between the straight classical trajectory and the ESL. Since $\mu\propto \sqrt{v}$, the value of the characteristic impact parameter decreases drastically once fields slow down.

The actual trapping event is complicated by Hubble damping as well as the fragmentation of fields once backreaction becomes important. Except in the simplest one-dimensional cases, a proper discussion of trapping requires numerical simulations, as in preheating \cite{Prokopec:1996rr,Felder:2000hq,Frolov:2008hy,Khlebnikov:1996zt,Khlebnikov:1996mc}. 

In this section, we take a binary point of view and assume that the moduli get trapped once they approach an ESL to within a distance smaller than $\mu$ (we go beyond this binary treatment in Sec.~\ref{sec:moduliareinflatons}). We further treat $\mu$ as a constant\footnote{We will find that $p_{trap}\propto \mu^{D-d-1}$; hence, trapping becomes less likely as $v$ decreases due to Hubble friction, so that we overestimate $p_{trap}$ by treating $\mu$ as a constant over the whole trajectory. A better estimate could be achieved by integrating a probability density over the trajectory $\vec{\varphi}(t)$ using $\mu(t)$. Since the latter is model dependent, we will be satisfied with the overestimate caused by treating $\mu\sim const$ (the correction usually introduces factors of order one).  }.

\subsection{Trapping Probability \label{sec:trappingprobability}}
 As we saw in section \ref{sec:length}, Hubble damping limits the overall path length to $s\lesssim few \times M_{pl}$ in an expanding universe. 
If the field trajectory $\vec{\varphi}(t)$ approaches an ESL to within the characteristic impact parameter $\mu$, trapping of the moduli is common, see Sec.~\ref{sec:trapping} and Fig.~\ref{pic:trapping}. We denote the dimensionality of the ESL by $d$ ($d\leq D$), so that $d=0$ corresponds to an ESP, $d=1$ to a line, etc.         

Further, we assume that ESLs are spread over moduli space with a characteristic distance $x$ (we make this more precise below).  For $x>\mu$, the probability of trapping is less than one,  provided that the $\varphi_i$ start out at a random initial position and velocity. Our goal is to estimate the dependence of this probability on the ratios $s/x$ and $\mu/x$,  as well as the scaling with the moduli space's and ESLs' dimensions, $D$ and $d$. Throughout this section, we take $s>x>\mu$ (see Sec.~\ref{sec:terminalv} for $x\sim \mu$).

\begin{figure*}[tb]
\includegraphics[scale=0.4,angle=0]{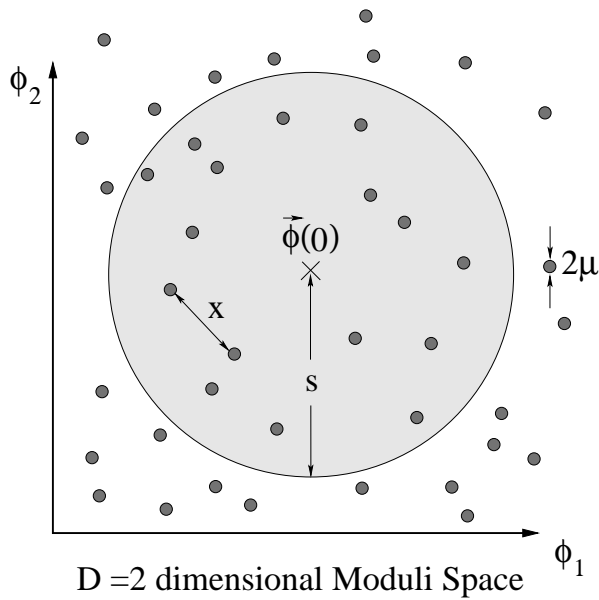}
   \caption{\label{pic:Dimensionalmoduli} Schematic of a higher dimensional moduli space (D=2 is shown): the large light-gray sphere of radius $s$ encloses the  accessible volume  by $\vec{\varphi}(t)$. The small dark-gray spheres are ESPs ($d=0$) of effective radius $\mu$, distributed with an average inter-ESP distance $x$. }
\end{figure*}

\subsubsection{Extra Species Points  (ESP, $d=0$)}
\begin{figure*}[tb]
\includegraphics[scale=0.4,angle=0]{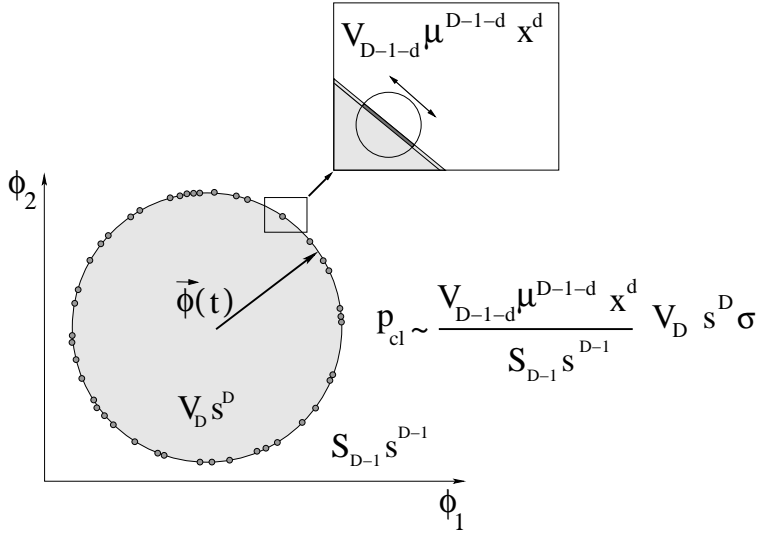}
   \caption{\label{pic:detail} Visualization of the trapping probability (the case $D=2$ and $d=0$ is depicted): we first push all ESLs (dark-gray) towards the boundary of the light-gray sphere of radius $s$. The ratio of the area that is blocked by the ESLs to the total surface area gives a lower bound of the trapping probability.}
\end{figure*}

We assume that ESPs are common and distributed over field space with a number density of $\sigma\equiv 1/x^D$. Consider a D-dimensional sphere of radius $s$ centered at $\vec\varphi(0)$, enclosing the region  that the moduli can access, see Fig.~\ref{pic:Dimensionalmoduli}. In this sphere we find
\begin{eqnarray}
n_{ESL}\equiv \sigma V_Ds^D = \frac{V_Ds^D}{x^D} \label{number}
\end{eqnarray}
ESPs that are potentially reachable by $\vec{\varphi}(t)$. 
 Here 
\begin{eqnarray}
V_D=\frac{\pi^{D/2}}{\Gamma\left(\frac{D}{2}+1\right)}
\end{eqnarray} 
is the volume of a D-dimensional unit-sphere, and $\Gamma$ is the Gamma-function ($\Gamma(x)\approx \sqrt{2\pi}x^{x-1/2}\exp(-x)$ for large $x$). Since the majority of the volume in a sphere is close to its surface for large $D$, the distance from $\vec{\varphi}(0)$ to the majority of the accessible ESPs is given by a distance close to $s$, the radius of the sphere. Thus, we approximate the distance of all ESPs to the center of the sphere by $s$. A lower bound for the probability of trapping can then be estimated by the ratio of the sphere's surface blocked by the ESPs to the total surface area
\begin{eqnarray}
p_{cl}\gtrsim \frac{n_{ESP} V_{D-1}\mu^{D-1}}{S_{D-1}s^{D-1}}\,, \label{probesp}
\end{eqnarray}  
where
\begin{eqnarray}
S_{D-1}=DV_D
\end{eqnarray}
is the surface area of a D-dimensional unit-sphere, see Fig.~\ref{pic:detail} for a schematic. Here, we assume that ESPs are sparse enough  to prevent overlapping of blocked regions on the sphere's surface so that our results are only valid for $p_{cl}\ll 1$. 
 Further, we take $D\geq 2$, since the one dimensional case is trivial: the probability of getting trapped is $p_{cl}=1$ if $x<s$ and $p_{cl}=s/x$ for $x>s$. (\ref{probesp}) is a lower bound, since even a few ESPs close to the center of the sphere, which we treated as if they where sitting on the surface, can block out a large fraction of the surface. We come back to this issue in appendix \ref{sec:volumeargument}.

With (\ref{number}) the lower bound on the probability reads
\begin{eqnarray}
p_{cl}\gtrsim \frac{1}{D}V_{D-1}\frac{s}{x}\left(\frac{\mu}{x}\right)^{D-1}\,. \label{probtarpesp}
\end{eqnarray}
In line with expectations based on \cite{Kofman:2004yc}, the more ESPs are within reach, the higher the probability of trapping. However, since $\mu<x$ and $p_{cl}\propto (\mu/x)^{D-1}$ we see that trapping becomes heavily suppressed for $D\geq 2$; thus, for large $D$, the starting point $\vec{\varphi}(0)$ needs to be within close range of an ESP, that is within a distance of order $\mu$ or less, to have a reasonable chance to get trapped. This should be compared to the one dimensional case, where the relevant distance is $s/x$: the modulus field is likely to be trapped as long as a single ESP is within reach.

The suppression $p_{cl}\propto (\mu/x)^{D-1}$ has important consequences for vacuum selection on the landscape. When the notion of quantum moduli trapping near ESPs was investigated in \cite{Kofman:2004yc,Watson:2004aq}, a dynamical selection principle to single out certain vacua associated with additional symmetries on the landscape, as used in \cite{Dine:2000ds,Dine:1998qr}, was a strong motivation. However, for large $D$ we see that $\vec{\varphi}$ needs to start out or be accurately aimed at an ESP in order to get trapped near it, hindering this dynamical selection. This argument remains true for loci of higher dimensions as long as $D\gg d$ and $x>\mu$.

\subsubsection{Extra Species Loci  (ESL, general $d$) \label{sec:classicalESL}}
There are many cases where symmetries are enhanced (and additional light states appear) not at points, but at loci of higher dimensionality. We can easily generalize the derivation of the last section to arbitrary $d$. We first assume a distribution of the ESLs on  moduli-space with a characteristic distance $x$, similar to the case of ESPs; for example, if $d=1$, one can visualize the loci as a string network with characteristic inter-string distance $x$. Thus, we can define an analog to the number density in the ESP case via $\sigma \equiv 1/x^{D}$, so that the blocked (by ESLs) surface area of the D-dimensional sphere with radius $s$ is bigger than $V_D s^D\sigma x^d\mu^{D-1-d}V_{D-1-d}$. Here we treated all loci as if they were located at  the surface of the sphere (hence, we find again a lower bound for the probability) and that they are sparse enough to ignore overlapping. As a consequence, the trapping probability satisfies
\begin{eqnarray}
p_{cl}\gtrsim \frac{1}{D}V_{D-1-d}\frac{s}{x}\left(\frac{\mu}{x}\right)^{D-1-d}\,. \label{trappingprobabilityclassicallb}
\end{eqnarray}     
For $d<D-1$ we observe a suppression of the trapping probability $\propto (\mu/x)^{D-1-d}$. Furthermore,  $V_{D-1-d}\rightarrow 0$ as $D-1-d\rightarrow \infty$,  providing  yet another suppression. See Appendix \ref{sec:examples} to see how this result applies to two simple examples.

 The factor of $1/D$ is absent if we estimate the probability more accurately by means of a volume argument, see Appendix \ref{sec:volumeargument}, which does not require any artificial shifting of ESLs,  yielding 
\begin{eqnarray}
p_{cl}\approx N_{\perp}V_{D-1-d}\frac{s}{x}\left(\frac{\mu}{x}\right)^{D-1-d}\,.\label{trappingprobabilityclassical}
\end{eqnarray}
Here, $N_{\perp}$ depends on the distributions of ESLs; for example, the grid-like distribution in Appendix \ref{sec:volumeargument} leads to (\ref{Nperp}),
\begin{eqnarray}
N_{\perp}=\left( {{D-1}\atop {d}}\right)=\frac{(D-1)!}{d!(D-1-d)!}\,.
\end{eqnarray} 
  
\subsection{Comments on Drifting Moduli \label{sec:driftingmoduli}}
During inflation any classical movement of moduli is quickly damped by Hubble friction, yet quantum mechanical fluctuations can still cause fields to drift through moduli space given that Hubble induced corrections to the moduli's masses are negligible and no classical potential is present. If we follow the field value in a given Hubble patch, we can observe a random walk. Hence, one might expect an increased trapping probability, since a random walk is more volume filling that a straight trajectory. Since the Hausdorff dimension of a random walk is $2$, that is a random walk is area filling, we expect the trapping probability to be slightly less suppressed, $p_{rw}\propto (\mu/x)^{D-2-d}$ instead of $p_{cl}\propto (\mu/x)^{D-1-d}$. This expectation is correct\footnote{Note that $\mu$ in appendix \ref{sec:trappingrandomwalk} has a different meaning: it is not related to the speed of the fields, but it is set by the distance at which the extra degrees of freedom can be described by a massless scalar in field theory. However, the scaling $p_{rw}\propto (\mu/x)^{D-2-d}$ is also valid for $\mu=\sqrt{v/g}$ if a classical trajectory is intertwined on small scales so that it can be approximated by a random walk in $D$ dimensions.}, see appendix \ref{sec:trappingrandomwalk}. We note that if the fields perform a random walk as described in appendix \ref{A:trappingrw}, and trapping does occur, then the trapping event has to take place before the last sixty e-folds of inflation; in that case, our entire Hubble patch today lies in a region that was inside the Hubble patch at the time of trapping and as a result, the moduli fields have the same value everywhere in our observable universe.

However, we are primarily interested in the limit $D\gg d$, leading to a strongly suppressed trapping probability if $\mu<x$. Then the mere presence of ESLs does not guarantee late time values of moduli fields near these locations; see appendix \ref{sec:expansioninflation} for more details.

In section \ref{sec:terminalv} we investigate the case $\mu \sim x$, in which case we expect particle production to be important; we further allow for a classical potential for the $\varphi_i$, so that they may be responsible for inflation.

\section{Moduli as Inflatons \label{sec:moduliareinflatons}}
So far, we focused on fields whose potential energy is negligible compared to both, their kinetic energy and $3 H^2 /(8\pi G )$ (from here on we set $8\pi G\equiv 1$). If the potential energy dominates, a phase of moduli-driven inflation results (see \cite{McAllister:2007bg}
 for a recent review on inflationary models in string theory); successful moduli trapping goes hand in hand with (p)reheating, while failed trapping events can still have interesting consequences. 

If moduli serve as inflatons, a successful trapping event is unwanted if it occurs during the last sixty e-folds of inflation. However, failed trapping events during inflation, that is some particle production at an ESL, slow down the fields and render steeper potential viable for slow-roll inflation. This is the idea behind trapped inflation \cite{Kofman:2004yc,Green:2009ds}, recently resurfaced as a special case of single field monodromy inflation \cite{Silverstein:2008sg} with sub-Planckian field excursions. Here, ESPs are repeatedly encountered by the inflaton as the angular modulus in a string compactification revolves around a periodic direction \cite{Silverstein:2008sg}, see also \cite{Brandenberger:2008kn}. Due to the classical (smooth) potential, the inflaton does not get trapped for long (even though $D=1$ and $d=0$) but it slows down until inflation dilutes the produced particles sufficiently and the field picks up speed again. After leaving the ESP, the produced $\chi$-particles quickly become heavy, re-scatter off the homogeneous inflaton condensate and cause a cascade of inflaton fluctuations in the IR \cite{Barnaby:2009mc,Barnaby:2009dd}. These fluctuations can dominate over the primordial ones and lead to observable signatures \cite{Barnaby:2009mc,Barnaby:2009dd}, such as a bump in the power-spectrum for each failed trapping event and potentially large non-Gaussianities.
 
One may wonder if it is conceivable to increase the number of fields in trapped inflation, for as long as each of the inflatons comes close to ESLs during inflation, the above mentioned effects will be present. 

\subsection{$x\gg\mu$, Suppressed Trapping}
Since the probability of getting close enough to ESLs depends on the type of movement through moduli space, we need to discern between three classes of inflation
\begin{enumerate}
\item Fields with a smooth potential.  A smooth classical trajectory in field space results, without any tunneling events. 
\item An overall smooth potential with small random bumps and dips. The classical trajectory can be described by a biased random walk; the overall drift is determined by the averaged potential, while the superimposed Brownian motion is caused by scattering events on the bumps, see i.e.~\cite{Tye:2008ef,Huang:2008jr,Tye:2009ff}. 
\item A potential with prominent valleys and mountains. The evolution of fields is the result of repeated tunneling/scattering events, not slow-roll, see i.e.~\cite{Tye:2008ef,Huang:2008jr}. 
 \end{enumerate}

If $D>d+1$ and fields roll slowly, as in case one, we expect trapping events to be rare, since $p_{cl} \propto (\mu/x)^{D-1-d}$ from (\ref{trappingprobabilityclassical}) and $\mu=\sqrt{v/g}$ from (\ref{defmu}).

If the trajectory is intertwined, as in case two, we need to discern between the drift velocity $\bar{v}$, which is again small, and the velocity along the trajectory $v=\left(\sum_i\dot{\varphi}_i^2\right)^{1/2}$. These two velocities are comparable if small scattering events dominate; thus, we expect again an absence of moduli trapping if $D>d+1$. However, if $v\gg \bar{v}$, for example due to frequent large angle scattering events, we note that $\mu$ is not suppressed. The trapping probability of the resulting biased random walk lies between the classical one in (\ref{trappingprobabilityclassical}) and  $p_{rw} \propto (\mu/x)^{D-2-d}$ from (\ref{probabilityrandomwalk}). To be concrete, if $\bar{v}N/H\gg s\sqrt{N n}$, where $s$ is the mean free path between two scattering events (not $H/2\pi$), $N$ is the number of e-folds and $n$ is the number of scattering events per e-fold, we may approximate $p_{brw}\approx p_{cl}$. On the other hand, if the classical drift is negligible, $\bar{v}N/H\ll \sqrt{N n}s$, we may use $p_{brw}\approx p_{rw}$.

In case three, the trajectory is well described by an unbiased random walk so that (\ref{probabilityrandomwalk}) is applicable. However, the step width $s$, number of steps per e-fold $n$, and $\mu$ need to be reconsidered, because the trajectory is the result of repeated tunneling/scattering events.

Regarding the viability of trapped inflation in the presence of multiple fields, we note that if $p_{cl}$ is applicable, there needs to be a series of closed $D-1$ dimensional ESL-hypersurfaces encompassing the origin to which $\vec{\varphi}(t)$ is heading; since $d=D-1$, the trapping probability (\ref{trappingprobabilityclassical}) is not suppressed, particle production can occur and trapped inflation may commence. If the fields perform a random walk, the dimensionality of ESLs can be smaller ($d=D-2$). Thus, trapped inflation, as envisioned in \cite{Kofman:2004yc,Green:2009ds} requires ESLs of dimensionality close to $D$.

However, as we shall see in Sec.\ref{sec:terminalv}, a compelling type of trapped inflation occurs naturally if $D$ is large and $\mu\sim x$ due to the presence of a terminal velocity, even if $d=0$. 

\subsection{$x\sim \mu$, a Terminal Velocity on the Landscape \label{sec:terminalv}}
Consider that the field-trajectory encounters a steep region on the landscape, unsuitable for slow-roll inflation. The classical potential leads to an increasing speed so that $v=|\dot{\vec{\varphi}}|$ becomes large. Such an increase in $v$ causes the effective trapping radius $\mu=\sqrt{v/g}$ to grow. For simplicity, we consider ESPs only, that is $d=0$ so that it is extremely unlikely to run head on into an ESP for large $D$ (a generalization to ESLs is straightforward as long as $D\gg d$). 

Once $\mu\sim x$, the trajectory is bound to come within range of neighboring ESPs. For example, if $D=1$, the inflaton becomes susceptible to $2$ ESPs, one ahead and one behind; if $D$ is large, the trajectory becomes within reach of many ESPs, of order $2^D$.

Around each of these ESPs, particles are produced (see Sec.~\ref{sec:particpleproductionwithimpactparameter} for the computation). As the inflatons continue to roll, there is an increasing number of particles in the wake of the trajectory \footnote{The particles are subsequently diluted due to the expansion of the universe and also back-scatter and decay, see Sec.~\ref{sec:particpleproductionwithimpactparameter}. }. As we shall derive below, the effect of each individual ESP is small and no actual trapping occurs if $D$ is large, but the combined backreaction of all produced particles leads to a resistive force opposite to $\dot{\vec{\varphi}}$. This situation is comparable to that of friction/viscosity in a fluid. 

For large $D$, this friction increases sharply once $\mu\sim x$, that is around
\begin{eqnarray}
v \sim g x^2\,,
\end{eqnarray}
preventing a further increase in $v$ and leading to a terminal velocity $v_t$ on the landscape, primarily determined by the density of ESPs and the coupling strength between the inflaton condensate and the new light degrees of freedom. Thus, depending on $x$ and $g$, inflatons could evolve slowly even if the potential is steep. Since no single ESP has a strong influence on $\vec{\varphi}$, we expect the trajectory to be largely unchanged, that is, $\vec{\dot{\varphi}}/v$ still follows the potential gradient, but with a maximal speed $v_{t}$ determined by balancing the drag force, not Hubble friction,  with the driving force of the potential. A similar argument can be made for ESLs as long as $d\ll D$. Note that this terminal velocity is present at weak coupling, opposite to the speed limit at strong coupling in \cite{Silverstein:2003hf} which led to DBI-inflation \cite{Alishahiha:2004eh}.

We conclude that driving inflation on the landscape may not require a flat potential after all. If ESLs are ubiquitous, fields evolve slowly due to quantum backreaction. This type of inflation is a generalization of one-dimensional trapped inflation, with the important difference that no actual trapping occurs. Furthermore, whereas ESPs need to be aligned neatly like pearls on a string if $D=1$, which may or may not be natural \cite{Silverstein:2008sg}, trapped inflation can result for randomly distributed ESPs in higher dimensions. Further, the terminal velocity becomes independent of the slope of the potential in the large $D$ limit.

In the following we analyze the physical processes behind higher dimensional trapped inflation in more detail, before providing a phenomenological model and computing some observational consequences.

\subsubsection{Particle Production near ESPs and Computation of the Terminal Velocity\label{sec:particpleproductionwithimpactparameter}}
Let us consider particle production at a single ESP which the trajectory passes by at a distance $\mu_0$ at $t_{\mbox{\tiny ESP}}$ with constant velocity $v$ (see Sec.\ref{sec:trapping} for the notation) \footnote{Since particle production near an ESP is exponentially suppressed the larger the distance to the ESP, the majority of particles are produced during a short time-frame, when the distance is minimal; this justifies treating $v$ as a constant while particles are produced.}. Ignoring backreaction onto the trajectory while $\vec{\phi}$ is close to the ESP, we can compute the number density of $\chi$ particles with wave-number $k$ after the encounter, $t>t_{\mbox{\tiny ESP}}$, to \cite{Kofman:1997yn,Kofman:2004yc}
\begin{eqnarray}
n_k=\exp\left({-\pi \frac{k^2+g^2\mu_0^2}{gv}}\right)\left(\frac{a(t_{\mbox{\tiny ESP}})}{a(t)}\right)^3\,,
\end{eqnarray}
where we incorporated the subsequent dilution for $t>t_{\mbox{\tiny ESP}}$ due to the expansion of the universe. The corresponding energy density is
\begin{eqnarray}
\rho_\chi=\int\frac{d^3k}{(2\pi)^3}n_k\sqrt{k^2+g^2(\vec{\varphi}(t)-\vec{\varphi}_{ESP})^2}\approx g\left|\vec{\varphi}(t)-\vec{\varphi}_{ESP}\right|n_{\chi}\,,
\end{eqnarray}
where $\vec{\varphi}_{ESP}$ is the location of the ESP, $\left|\vec{\varphi}(t)-\vec{\varphi}_{ESP}\right|=\sqrt{\mu_0^2+v^2(t-t_{\mbox{\tiny ESP}})^2}$, and
\begin{eqnarray}
n_{\chi}=\int\frac{d^3k}{(2\pi)^3}n_k\approx \frac{(gv)^{3/2}}{(2\pi)^3}e^{-\pi g \mu_0^2/v}
\left(\frac{a(t_{\mbox{\tiny ESP}})}{a(t)}\right)^3\,. \label{nchi}
\end{eqnarray}
We assume that $\chi$-particles are sufficiently stable to ignore their decay, but the unavoidable scattering off the inflaton condensate lowers the number density over time by an additional factor of at most $\exp({-g^2n_\chi/(2\mu_0v))} \leq 1$ \cite{Kofman:2004yc}. For simplicity, we ignore this factor in the following \footnote{The presence of a terminal velocity and even its value in the large $D$ limit does not change if this factor is kept.}. 

Before we compute the backreaction onto $\vec{\varphi}$, let us investigate up to which distance we need to consider ESPs. If we find ourself at a random position in field space, one might guess that only nearest neighbors roughly a distance $x$ away need to be considered. However, there are two effects which render the effective impact parameter $\mu_0$ bigger than $x$. Consider a D-dimensional sphere of radius $l$ in field space with an ESP density $\sigma=1/x^D$; solving  $V_Dl^D\sigma=1$ gives the radius above which we expect at least one ESP inside the sphere. Since $V_D=\pi^{D/2}/\Gamma(D/2+1)$ and using the asymptotic form of the Gamma function for large arguments, $\Gamma(x)\approx \sqrt{2\pi}x^{x-1/2}\exp(-x)$, we arrive at 
\begin{eqnarray}
l\approx %x\sqrt{\frac{D+2}{2\pi e}}\approx 
x\sqrt{\frac{D}{2\pi e}}\,,
\end{eqnarray}
 considerably larger than $x$ for $D\gg 1$. 
 
\begin{figure*}[tb]
\includegraphics[scale=0.4,angle=0]{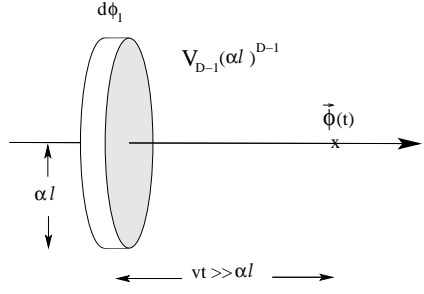}
   \caption{\label{pic:save} The energy density $\rho_\chi^{\mbox{\tiny total}}$ is dominated by $\chi$-particles in the proximity of ESPs near the boundary of a cylinder of radius $\mu_0=\alpha l$ around the classical trajectory $\vec{\varphi}(t)$.}
\end{figure*}

 The effect of ESPs that are further away is exponentially suppressed according to (\ref{nchi}), but their increased number can counterbalance this suppression to some degree. As a consequence, we need to consider not only nearest neighbor ESPs. Let us compare the energy density of $\chi$-particles at ESPs that were produced at $t_{\mbox{\tiny ESP}}$ up to a distance of $\alpha l$ perpendicular to the trajectory; we are interested in their energy density when $(t-t_{\mbox{\tiny ESP}})v\gg \alpha l$, see Fig.~\ref{pic:save}. Noting that for large $D$ almost all ESPs are separated from $\vec{\varphi}(t_{\mbox{\tiny ESP}})$ perpendicular to the trajectory and that the majority of ESPs can be found near the rim of the cylinder with radius $\alpha l$ in Fig.~\ref{pic:save}, we can approximate the ratio of the $\chi$-particles' energy density  near the rim to the one of particles at nearest neighbor ESPs by
\begin{eqnarray}
\frac{\rho_{\mbox{\tiny tot}}^{\mbox{\tiny rim}}}{\rho_{\mbox{\tiny tot}}^{\mbox{\tiny nearest}}}\approx \alpha^{D-1}\exp\left(-{\frac{\pi g l^2(\alpha^2-1)}{v}}\right)\,.
\end{eqnarray}
This ratio is maximal for 
\begin{eqnarray}
\alpha = \Bigg\lbrace  \begin{array}{l}
			1\,,\,\,\mbox{if}\,\, v<\frac{gx^2}{e}\,, \\
			\sqrt{\frac{e}{gx^2}v}\,,\,\,\mbox{if}\,\, v\geq \frac{gx^2}{e}\,, \label{alpha}
		 \end{array}
\end{eqnarray}
where we ignored terms of order $1/D$ and used $l^2\approx x^2D/(2\pi e)$. Thus, the energy density is dominated by $\chi$-particles a distance 
\begin{eqnarray}
\mu_0=\alpha l \label{mu0trapped}
\end{eqnarray}
 away, which sets the characteristic impact parameter for the majority of relevant ESPs. 
 
Next, we compute the backreaction onto the inflationary trajectory in the Hartree approximation \cite{Traschen:1990sw,Kofman:1997yn,Kofman:2004yc}. We first note that the combined effect of ESPs in the directions perpendicular to the trajectory becomes exceedingly suppressed with increasing $D$: since $\mu_0\propto \sqrt{D}$, the distribution of ESPs near the rim of the cylinder in Fig.~\ref{pic:save} can be well approximated by a continuous ESP density; this approximation is similar to treating electric charges distributed on a macroscopic metal cylinder as a continuous surface charge density. Then, due to the symmetry of the configuration, the net force perpendicular to the trajectory vanishes, but a resistive force opposite to $\dot{\vec{\varphi}}$ remains. We assume that the potential is such that the trajectory is straight over times-scales of interest \footnote{If $|\dot{\vec{\varphi}}|=v_t$ during the last sixty e-folds of inflation, the COBE normalization imposes $H\approx 0.018 \sqrt{v_t}$, see forward to (\ref{inflationaryscale}). We deduce that the trajectory needs to be straight over $\Delta \varphi\gtrsim 19 \sqrt{v_t}$, which should be sub-Planckian. Hence, the terminal velocity needs to be
\begin{eqnarray} 
v_t\lesssim 10^{-4} \label{footnote:straight}
\end{eqnarray}
for internal consistency.}, that is over $\Delta t \gtrsim 1 / (3H)$ so that
 $(a(t-\Delta t)/a(t))^3 \lesssim 1/e$. 
Denoting with $\varphi_1$ the field along the trajectory, the equation of motion for $\varphi_1$ becomes
\begin{eqnarray}
\ddot{\varphi}_1+3H\dot{\varphi_1}+\frac{\partial V}{\partial \varphi_1}+g^2\left<\chi\right>^2\varphi_{1}=0\,, \label{EOMphi1}
\end{eqnarray}
where 
\begin{eqnarray}
\left|g^2\left<\chi\right>^2\varphi_{1}\right|&\approx & \int_{\mbox{\tiny traj.}} g n_{\chi}V_{D-1}\mu_0^{D-1}x^{-D} d\varphi_1\\
&\approx&\frac{(gv)^{5/2}}{(2\pi)^3}\frac{1}{3H}\exp\left(-\pi g\mu_0^2/v\right)V_{D-1}\frac{\mu_0^{D-1}}{x^D}
\end{eqnarray}
with $\mu_0$ from (\ref{mu0trapped}) and we took $v\approx \mbox{const}$ as well as $H\approx \mbox{const}$, so that $\int_{t_{\mbox{\tiny ESP}}}^t (a(t_{\mbox{\tiny ESP}})/a(t))^3 dt_{\mbox{\tiny ESP}}\approx 1/{3H}$. For $v<\frac{gx^2}{e}$ there is a strong exponential suppression $\propto \exp(-Dgx^2/(2ev))$, so that backreaction is negligible and trapped inflation is impossible, in agreement with our prior heuristic arguments. We recognize
\begin{eqnarray}
v_c\equiv \frac{gx^2}{e}
\end{eqnarray}
as the aforementioned critical velocity of order $gx^2$ above which backreaction from particles near ESPs becomes important. 
In the regime $v\geq v_c$ the drag increases sharply: assuming\footnote{Using the COBE normalization, see forward to (\ref{inflationaryscale}), yields  $H\approx 0.018 \sqrt{v_t}$ and the condition $3H|\dot{\varphi}_1|\ll |\partial V/ \partial \varphi_{1}|$ imposes an upper bound on the terminal velocity of 
\begin{eqnarray}
v_t^{3/2}\ll 18.5 |\partial V/ \partial \varphi_{1}|\,,\label{req1}
\end{eqnarray} 
indicating that the potential needs to be steep enough for $v_t$ to provide a speed limit. In essence, this is the opposite of the requirement for a potential to support slow-roll inflation.}
$\left|\ddot{\varphi}_1 \right|\ll 3H|\dot{\varphi}_1|\ll |\partial V/ \partial \varphi_{1}|$ in (\ref{EOMphi1}), we can compute the terminal velocity by equating the drag with the driving force caused by the potential gradient to
\begin{eqnarray}
v_t=|\dot{\varphi}_1|\approx gx^2\Delta
%v_c^{(D-1)/(D+4)} e^{D/(D+4)} \Delta  
\label{terminalvelocity}
\end{eqnarray}
where we defined
\begin{eqnarray}
\Delta\equiv \left(\frac{(2\pi)^33H}{g^{5}x^4} \frac{\partial V}{\partial \varphi_1} \right)^{2/(D+4)}\,.
%\left(\frac{(2\pi)^33Hx}{g^{5/2}}\sqrt{\frac{D}{2\pi e}}\left|\frac{\partial V}{\partial \varphi_1}\right|\right)^{2/(D+4)}\,.
\end{eqnarray}
Note that $\Delta\rightarrow 1$ in the limit of $D\rightarrow \infty$; as a consequence, the terminal velocity becomes independent of the slope of the potential. Further, keeping only the leading order terms in $1/D$, we recognize that the terminal velocity is simply
\begin{eqnarray}
v_t\approx gx^2
\end{eqnarray}   
in the large $D$ limit, in agreement with our heuristic arguments of section \ref{sec:terminalv}. This is a remarkably simple result, providing a general speed limit on the landscape entirely determined by the distribution of ESPs and the coupling strength of the additional light degrees of freedom to the moduli, but independent of the slope of the potential.

If the classical trajectory is intertwined on small scales ($\Delta \varphi\lesssim 19 \sqrt{v_t}$) the above derivation is inapplicable; we comment on this regime in Sec.~\ref{sec:meander}.

\subsubsection{Different Regimes of Trapped Inflation \label{sec:regimesoftrappedinflation}}
Given a reasonably straight classical trajectory, there are two distinct regimes of inflation that could occur once $v$ becomes of order $v_t$:
\begin{enumerate}
\item The classical trajectory is followed with a constant velocity close to $v_t$; this requires a large $D$ to prevent overshooting and extended oscillations around the terminal velocity.
\item The classical trajectory is followed with a velocity that oscillates around $v_t$, never settling into the constant velocity regime before inflation ends.
\end{enumerate} 
To decide which type of inflation takes place, let us estimate the relevant time scale at which backreaction of particles near ESPs becomes stronger than the classical force due to the potential. In this section we neglect the expansion of the universe; as the velocity approaches $v_t$ from below we expand 
\begin{eqnarray}
v\approx v_t+a_1 t
\end{eqnarray}
before backreaction becomes important; here, we defined $a_1\equiv\left|\partial V/\partial \varphi_1\right|_{t_{\mbox{\tiny ESP}}}$ and set $t_{\mbox{\tiny ESP}}=0$ (particles start to be produced at $t=0$). Hence, the characteristic time-scale for the classical force is
\begin{eqnarray}
t_{\mbox{\tiny cl.}}\equiv \frac{v_t}{\left|\partial V/\partial \varphi_1\right|_{t_{\mbox{\tiny ESP}}}}\,.
\end{eqnarray}
On the other hand, the force due to backreaction increases in time as
\begin{eqnarray}
\left|g^2\left<\chi\right>^2\varphi_{i}\right|&\approx & \int_{\varphi_1(0)}^{\varphi_1(t)} g n_{\chi}V_{D-1}\mu_0^{D-1}x^{-D} d\varphi_1\\
&\approx&\frac{(g(v_t+a_1 t))^{5/2}}{(2\pi)^3}e^{-D/2}\left(\frac{e(v_t+a_1 t)}{gx^2}\right)^{(D-1)/2}\sqrt{\frac{2\pi e}{D}}\sqrt{\frac{D+1}{2\pi}}\frac{1}{x} t \\
&=&\frac{(gv_t)^{5/2}}{(2\pi)^3}\left(1+\frac{t}{t_{\mbox{\tiny cl.}}}\right)^{(D+4)/2}\sqrt{\frac{D+1}{D}}\frac{t}{x}\,,
\end{eqnarray}
where we used $V_{D-1}\mu_0^{D-1}/x^D\approx \alpha^D/\mu_0 \sqrt{(D+1)/(2\pi)}$, $\alpha=\sqrt{ev/(gx^2)}$ from (\ref{alpha}) and $\mu_0$ from (\ref{mu0trapped}). Setting $\left|g^2\left<\chi\right>^2\varphi_{i}\right|=\left|\partial V/\partial \varphi_1\right|_{t_{\mbox{\tiny ESP}}}$ and solving for $t$ leads to the characteristic time-scale for backreaction to dominate over the classical force,
\begin{eqnarray}
t_{\mbox{\tiny backr.}}&\approx& t_{\mbox{\tiny cl.}} \frac{2}{D+4} \mbox{LambertW}\left(\frac{D+4}{2}\left|\frac{\partial V}{\partial\varphi_1}\right|_{t_{\mbox{\tiny ESP}}}^2\frac{x}{v_t}\sqrt{\frac{D}{D+1}}\frac{(2\pi)^3}{(gv_t)^{5/2}}\right) \label{tbackreaction}\\
&\approx& t_{\mbox{\tiny cl.}}
\frac{2}{D}
\ln\left(\frac{D}{2} \left| \frac{\partial V}{\partial\varphi_1} \right|_{t_{\mbox{\tiny ESP}}}^2 \frac{(2\pi)^3}{(gx)^6}\right)
\end{eqnarray}
Here, we used $(\dots)^{2/(D+4)}\approx 1+\ln(\dots)2/(D+4)$ before solving for $t_{\mbox{\tiny backr.}}$, and in the last step we kept leading order terms in $1/D$ only and used $v_t=gx^2$ as well as the asymptotic form of the LambertW function for large arguments, $\mbox{LambertW}(x)\approx \ln(x)$. For
\begin{eqnarray}
t_{\mbox{\tiny backr.}}\ll t_{\mbox{\tiny cl.}} \label{conditiononD}
\end{eqnarray}
 backreaction dominates before the classical potential has a chance to significantly alter the velocity. As a consequence, the classical trajectory is followed with a constant velocity close to $v_t$ once $v$ reaches $v_t$, without prolonged oscillations around $v_t$. 
(\ref{conditiononD}) along with (\ref{tbackreaction}) imposes a lower bound on the dimensionality of field space, 
$D\gg 2\ln(|\partial V/\partial \varphi_1|_{t_{\mbox{\tiny ESP}}}^2(2\pi)^3/(gx)^6 )$.
If we take a steep potential $\left|\partial V/\partial \varphi_1\right|_{t_{\mbox{\tiny ESP}}}\sim 1$, large\footnote{Constraints from IR cascading require $g\lesssim 0.2$, see Sec.\ref{sec:IR-cascading}.} 
couplings $g\sim 0.1$  and ubiquitous ESPs $x \lesssim 10^{-2}$ so that $v_t\lesssim 10^{-4}$ and (\ref{footnote:straight}) is satisfied, we get  $D\gg 94$. 
If the potential happens to be shallower, as in Sec.\ref{sec:trappedinfllargeD}, lower values of $D$ suffice, whereas smaller values of $gx$ require slightly bigger values of $D$. Note that regardless of the slope, we need at least $D>10$, since we ignored terms of order $1/D$ throughout.

If $D$ is less then the minimal value imposed by (\ref{conditiononD}), one might still get inflation of the second type: the velocity overshoots, even more light particles are produced since $\mu_0$ increases, backreaction sets in and the velocity is forced\footnote{The velocity may actually turn around so that the trajectory becomes temporarily trapped near $\vec{\varphi}(t_{\mbox{\tiny ESP}})$ up until $\chi$ particles are diluted sufficiently by the expansion of the universe and $v$ increases again.} again well below $v_t$, mimicking the speed of cars during rush hour. Due to the changing velocity, we expect bumps in the power-spectrum as well as strong additional non-Gaussianities, rendering this regime less appealing. We leave this class of models to future studies.

\subsection{Phenomenology of Trapped Inflation in Higher Dimensions, $v\sim \mbox{const}$ \label{sec:trappedinfllargeD}}

Consider a simple multi-field inflationary model, $V=\sum_i V_i$ with $V_i=m^2\varphi_i^2/2$
and $\varphi_i(t_{ini})=\varphi_j(t_{ini})$ for all $i,j=1\dots D$. If we denote the effective field along the trajectory with $\varphi\equiv (\sum_i\varphi_i^2)^{1/2}$, we get $V(\varphi)=m^2\varphi^2/2$. Let us consider a regime that does not yield slow-roll inflation, $\varphi_{ini}\equiv \varphi(t_{ini})= 1$ so that field values remain sub-Planckian (we use $8\pi G \equiv 1$ if not stated otherwise in this section). If the trajectory is in a $D$-dimensional space ($D\gg 1$) with ubiquitous ESPs (or more generally ESLs), we concluded in section \ref{sec:particpleproductionwithimpactparameter} that the speed of $\varphi$ is bounded from above by the terminal velocity $v_{t}$, see equation (\ref{terminalvelocity}). Furthermore, in the large $D$ limit $v_t\approx gx^2$, independent of the slope of the potential; as a working example that satisfies all constraints in this paper, we choose $g=0.1$ and $x=10^{-3.5}$; then the coupling constant is below the bound originating from IR-cascading (see forward to Sec.~\ref{sec:IR-cascading}), $v_t=10^{-8}$ is well below the bound in (\ref{footnote:straight}) originating from the requirement that the distance over which the trajectory has to be straight be sub-Planckian, inflation lasts long enough,  see forward to (\ref{enoughefolds}), and (\ref{req1}) is satisfied so that Hubble friction is subdominant. The dimensionality of moduli space needs to be above the bound imposed by (\ref{conditiononD}) to guarantee that $\varphi$ rolls at the terminal velocity \footnote{Since the COBE normalization sets the inflationary scale in (\ref{inflationaryscale}), the mass becomes $m\approx\sqrt{6}\times 0.018 \sqrt{v_{t}}$ for $\varphi_{ini}\sim 1$; then the bound on $D$ imposed by (\ref{conditiononD}) becomes $D\gg 37$ since $\partial V/\partial \varphi=m^2\varphi_{ini}\approx 0.0019 v_t$. Note that the slope satisfies  $(\partial V/\partial \varphi)/(3H)\approx m\sqrt{(2/3)}>v_{t}$ so that the terminal velocity is indeed a speed limit. One can further check that 
\begin{eqnarray}
\rho_\chi\ll v_t^2/2\ll \rho_\varphi\,,\label{energyrequirement}
\end{eqnarray}
 as required for inflation to take place ($\chi$-particles dominate the equation of motion for $\varphi$, but their energy remains subdominant.).}, but the value of $D$ does not enter directly into observables as long as it is large enough ($D\gtrsim 37$ is sufficient in this section). The Hubble slow evolution parameter becomes 
\begin{eqnarray}
\varepsilon&\equiv&-\frac{\dot{H}}{H^2}
=\frac{3v_{t}^2}{2V}
\approx \frac{1}{6.5\times 10^{-4}}v_t\,,
\end{eqnarray}
where we used the COBE normalization in (\ref{inflationaryscale}) to set the inflationary scale. Since $\varepsilon$ should be much smaller than one, we find an upper bound for the terminal velocity 
\begin{eqnarray}
v_t\lesssim 10^{-4}\,,
\end{eqnarray}
which coincides with the one in (\ref{footnote:straight}). Inflation takes place up until $\epsilon\sim 1$, that is for $\varphi>\varphi_{end}\sim \sqrt{3}v_t/m\approx 0.004 \ll 1$. The energy density at this instant is $\rho=2v_t^2=2\times 10^{-14}$, corresponding to an upper bound on the reheating temperature\footnote{The actual reheating temperature depends on several factors: if preheating via parametric resonances (unlikely, see Sec:\ref{sec:examplepreheating}) and/or tachyonic instabilities takes place, the reheating temperature is close to this upper bound, which puts it at odds with bounds originating from thermal relics, such as gravitinos ($T_r\lesssim (10^6-10^8)\,\mbox{GeV}$, see \cite{Ellis:1984eq,Khlopov:1984pf,Kawasaki:2008qe} or \cite{Battefeld:2009sb} for a brief review).
 If reheating takes place via the old theory of reheating \cite{Dolgov:1982th,Abbott:1982hn,Kofman:1997yn}, the inflatons continue to oscillate around $\vec{\varphi}=0$  until $H$ becomes comparable to their decay rate(s) $\Gamma$. 
An estimate of the reheating temperature is then given by $T_r\approx \left( 90\Gamma^2/(8\pi^3 N_r G) \right)^{1/4}$.} of $T_r\lesssim (30 \rho/(8\pi^3N_rG))^{1/4}\sim 8.8\times 10^{11}\,\mbox{GeV}$ where we used $N_r\sim 1000$ as a estimate for the number of relativistic degrees of freedom \cite{Kofman:1997yn} and reinstated $8\pi G$. To put this another way, to guarantee a high enough temperature for nucleosynthesis, say $T>10\,\mbox{MeV}$, we need a terminal velocity of at least 
\begin{eqnarray}
v_t>10^{-35}\,, \label{reqreheating}
\end{eqnarray}
 which is a mild lower bound on $v_t$.

The number of e-folds becomes
\begin{eqnarray}
N&=&\int_{t_{end}}^{t_{ini}} H\, dt=\int^{\varphi_{ini}}_{\varphi_{end}} \frac{H}{\dot{\varphi}}\, d\varphi\approx\frac{1}{v_{t}}\frac{m}{2\sqrt{6}}\varphi_{ini}^2\,, \label{efolds}
\end{eqnarray}
where we neglected $\varphi_{end}\ll \varphi_{ini}$. As a consequence, we can write
\begin{eqnarray}
\varepsilon\simeq \frac{\varphi_{ini}^2}{8N^2}\,. \label{varepsilon}
\end{eqnarray}
Using the inflationary Hubble scale from (\ref{inflationaryscale}) and $\varphi_{ini}= 1$ in (\ref{efolds}), we get the so far strongest upper bound on the terminal velocity
\begin{eqnarray} 
v_t\lesssim 10^{-8} \label{enoughefolds}
\end{eqnarray}
from demanding that $N\gtrsim 60$ in order to solve the usual problems (flatness, horizon, etc.) of the big bang. We chose the inter-ESP distance $x=10^{-3.5}$ in hindsight to satisfy this observational constraint. For different potentials, the numerical pre-factor in (\ref{varepsilon}) changes, but the scaling $\varepsilon\propto \varphi_{ini}^2/N^2$ remains. Note that $\varepsilon$ is smaller than in the slow-roll case where $\varepsilon_{SR}\propto \varphi_{ini}^2/N$. 

\subsubsection{Observational Consequences \label{sec:obscons}}
Let us focus on adiabatic perturbations, which are properly described by (the Fourier modes of) the Mukhanov variable $v_k=z\zeta_k$ with $z=a\dot{\varphi}/H$ (see \cite{Bassett:2005xm} for a review). Here $\zeta$ is the curvature perturbation on uniform density surfaces. We are interested in evaluating its power-spectrum, which gets imprinted onto fluctuations of the cosmic microwave background radiation after inflation. If $\varepsilon$ evolves slowly, which is the case during inflation, we can approximate
\begin{eqnarray}
a(\tau)\propto (-\tau)^{-(1+\varepsilon)}\,, \label{bgra}
\end{eqnarray}
where $\tau=-\infty \dots 0$ is conformal time, $a\, d\tau=dt$ and $\partial (\,\,)/\partial \tau=(\,\,)^\prime$, solve the equation of motion for $v_k$
\begin{eqnarray}
v_k^{\prime\prime}+\left(k^2-\frac{z^{\prime\prime}}{z}\right)v_k=0
\end{eqnarray}
analytically in terms of Hankel functions \cite{Mukhanov:1990me,Bassett:2005xm}, match to the Bunch Davies vacuum in the far past ($v_k=\exp(-ik\tau)/\sqrt{2k}$), expand the solution on large scales and translate back to $\zeta_k$. 

In the end, the amplitude of the power-spectrum at horizon crossing ($k=aH$) becomes
\begin{eqnarray}
\mathcal{P}_\zeta=\left(\frac{H^2}{2\pi\dot{\varphi}}\right)^2_{k=aH}
\end{eqnarray}
to leading order in $\epsilon$, which is set by the COBE normalization $P_{\zeta}=(2.41\pm 0.11) \times 10^{-9}$ \cite{Komatsu:2008hk}. This means we need an inflationary scale of 
\begin{eqnarray}
H_{\mbox{\tiny inf.}}\approx 0.018 \sqrt{v_t}\,.\label{inflationaryscale}
\end{eqnarray} 
Further, the scalar spectral index reads
\begin{eqnarray}
n_s-1&=&\frac{d\,\ln \mathcal{P}_{\zeta}}{d\, \ln k}=3-2\nu\\
&\simeq & -4\varepsilon \label{scalarspectralindex}\approx -\frac{\varphi_{ini}^2}{2N^2} \,, 
\end{eqnarray}
where $\nu$ is defined by
\begin{eqnarray}
\frac{z^{\prime\prime}}{z}\equiv \frac{\nu^2-\frac{1}{4}}{\tau^2}
\end{eqnarray}
and we used $z^{\prime\prime}/z\simeq 2(1+3\varepsilon)/\tau^2$. The power-spectrum is closer to scale invariance than in the slow-roll case, since $\varepsilon\propto \varphi_{ini}^2/N^2$.

The power-spectrum of gravitational waves has an amplitude and spectral index of \cite{Bassett:2005xm}
\begin{eqnarray}
\mathcal{P}_{T}=\frac{2H^2}{\pi^2}\,\\
n_T=-2\varepsilon\,,
\end{eqnarray}
which leads to the usual expression for the tensor to scalar ratio $r$ 
\begin{eqnarray}
r\equiv \frac{\mathcal{P}_T}{\mathcal{P}_\zeta}=16\varepsilon\,.
\end{eqnarray}
The small value of $\varepsilon$ renders $r$ unobservably small with the current generation of experiments, such as Planck \cite{Planck}.

\subsubsection{Discussion}
In the simplest realization of trapped inflation in higher dimensions with $v=v_t=\mbox{const}$, the Hubble slow evolution parameter is suppressed by an additional power of $N$ (the number of e-folds) if compared to a slow-roll setup. As a consequence, the scalar power-spectrum is closer to a scale invariant one, in tension with the WMAP5 results \cite{Komatsu:2008hk} at the $2\sigma$ level, and the tensor to scalar ratio $r$ is unobservably small. Thus, one could firmly rule out this model of trapped inflation in higher dimensions if error bars around $n_s<1$ tighten or primordial gravitational waves of order $r\sim 10^{-2}$ be observed. 

However, we worked with a simple potential resulting in a straight trajectory. In general, one might not expect such a smooth potential on the landscape, but one leading to a curved trajectory. In such a case we can not neglect isocurvature perturbations, since they feed into the adiabatic mode with every twist and turn, leading to additional deviations from scale invariance and potentially observable non-Gaussianities. Further, if the trajectory twists and turns on small scales, our derivation of the terminal velocity is not applicable any more.
We comment on this regime in Sec.\ref{sec:meander}.

We would like to emphasize that the potential does not influence the speed of the effective field along the trajectory in trapped inflation if $D$ is large. Hence, less tuning of the potential is needed, but we require a high dimensionality of field space (so that backreaction increases sharply at $v_t$ and (\ref{conditiononD}) is satisfied) and ubiquitous ESPs/ESLs ($x\lesssim 10^{-3.5}$, so that $v_t$ is sufficiently small, and inflation lasts long enough, see (\ref{enoughefolds})). Further, we impose $g\lesssim 0.2$ to prevent perturbations from IR-cascading to dominate, see Sec.~\ref{sec:IR-cascading}. Other conditions imposed throughout are
\begin{itemize}
\item the trajectory should not be strongly curved over a distance covered in a few Hubble times, (\ref{footnote:straight}),
\item the slope of the potential needs to be large enough to guarantee that Hubble friction can be ignored, (\ref{req1}),
\item the potential energy needs to dominate over the kinetic one, which in turn needs to dominate over $\rho_\chi$, to guarantee that the universe inflates (\ref{energyrequirement}),
\item the reheating temperature needs to be large enough for nucleosynthesis to take place, (\ref{reqreheating}).
\end{itemize}
These requirements are all satisfied in the simple model presented in Sec.~\ref{sec:trappedinfllargeD}.
 
Another advantage of trapped inflation over ordinary slow-roll inflation is its independence on the initial velocity, a quantity that requires fine tuning in many slow-roll models. If the initial velocity happens to be bigger than $v_t$, a plethora of light states is produced rapidly whose backreaction causes a sudden decrease of $v$ well below $v_t$, potentially leading to temporary trapping near the initial location. Once $\chi$-particles are diluted by the expansion of the universe, the speed picks up again due to the classical potential and settles into the constant velocity regime of trapped inflation if $D$ is large enough.

\subsection{Additional Observational Signatures}
We would like to comment on two additional observational signatures brought forth by particle production during inflation.
\subsubsection{IR-Cascading \label{sec:IR-cascading}}
A crucial aspect of trapped inflation is the ongoing production of particles during inflation near ESPs.  As $\chi$-particles become heavy, they re-scatter off the homogeneous inflaton condensate and cause a cascade of inflaton fluctuations in the IR \cite{Barnaby:2009mc,Barnaby:2009dd}, leading to observable signatures in the power-spectrum. For particle production at a single ESP, an additional bump like feature at $k_{\mbox{\tiny ESP}}\sim H$ results, which can be approximated by \cite{Barnaby:2009mc,Barnaby:2009dd}
\begin{eqnarray}
\mathcal{P}_{\mbox{\tiny ESP}}\approx A_{\mbox{\tiny ESP}} \left(\frac{\pi e}{3}\right)^{3/2}\left(\frac{k}{k_{\mbox{\tiny ESP}}}\right)^3e^{-\frac{\pi}{2}\left(\frac{k}{k_{\mbox{\tiny ESP}}}\right)^2} \label{shapebump}
\end{eqnarray}
where the amplitude scales with the coupling $g$; for example, comparison with lattice field theory in a simple, one dimensional $m^2\varphi^2$ inflationary model yields $A_{\mbox{\tiny ESP}}\simeq 1.01\times 10^{-6}g^{15/4}$ for $g^2=1,0.1,0.01$. Note that the shape in (\ref{shapebump}) is only reliable for $10^{-7}\simeq g^2 \simeq 1$. Bumps with an amplitude of up to $10\%$ of the primordial power-spectrum on scales probed by current CMBR experiments are still consistent with observations \cite{Barnaby:2009dd}, leading to an upper limit on the coupling of $g^2\lesssim 10^{-2}$.

However, in the case of trapped inflation as investigated in Sec.\ref{sec:trappedinfllargeD}, particles are produced at a multitude of ESPs. The resulting superposition of bumps leads to a constant contribution to the power-spectrum. It is expected that this contribution, albeit scale invariant, is strongly non-Gaussian and thus ruled out as the dominant one \cite{Barnaby:2009dd} \footnote{A computation of Non-Gaussianities from IR cascading is in preparation by the authors of \cite{Barnaby:2009dd}; observational bounds on non-Gaussianities might yield stronger constraints on particle production during inflation than considerations based on the power-spectrum.}. 

In \cite{Barnaby:2009dd} observational bounds on the amplitude of features were given under the assumption that all ESPs lie on the trajectory, as in one dimensional monodromy inflation \cite{Silverstein:2008sg}, resulting in a superposition of bumps with identical amplitudes. For a large ESP or bump density ($k_i=k_1e^{(i-1)\Delta}$ with $\Delta \ll 1$), it was found that the constant contribution scales as $\Delta^{-1}$; consequently, the maximally allowed amplitude is proportional to $\Delta$, roughly $A\lesssim 2\Delta \times 10^{-9}$, which provides and upper bound on $g^2$.  

This may sound alarming, since trapped inflation in higher dimensions relies on a multitude of ESPs, roughly $n_{\mbox{\tiny ESP}}=V_{D}\mu_{0}^D/x^D\approx \alpha^D\approx e^{D/2}\gg 1$, at any given time. There is however a crucial difference, rendering the conclusions of \cite{Barnaby:2009dd} only indirectly applicable. The $\chi$-particles that dominate the energy density $\rho_{\chi}$ in Sec~\ref{sec:particpleproductionwithimpactparameter} are produced at ESPs a distance $\mu_0=\sqrt{D/(2\pi)}x$ away from the trajectory. As a consequence, particle production at any single ESP is weak; to be concrete, the particle number at each ESP is exponentially suppressed 
\begin{eqnarray}
\nonumber n_k&=&\exp\left(-\pi\frac{k^2+g^2\mu_0^2}{gv_t}\right)\\
&\approx& \exp\left(-\frac{\pi k^2}{g^2x^2}\right) \exp\left(-\frac{D}{2} \right)\,.
\end{eqnarray}
The amplitude of each bump should carry the same exponential suppression $\propto e^{-D/2}$.  Hence, the effect of $n_{\mbox{\tiny ESP}}\approx e^{D/2}$ ESPs at any given time leads to a constant contribution to the power-spectrum with a magnitude corresponding to the amplitude of a single feature, $A\simeq 10^{-6}g^{15/4}$; then the upper bound of order $A\lesssim 2\times 10^{-9}$ yields the relatively weak bound of $g\lesssim 0.2$. Of course, this heuristic argument should be followed up with a more rigorous computation, which we leave to a future publication.

We would like to point out that large values of $g$ are not required for trapped inflation in higher dimensions to operate; in fact, lower values of $g$ might be desirable, since smaller terminal velocities can be achieved with less ubiquitous ESPs.  

\subsubsection{Individual Massive $\chi$-particles\label{sec:massive}} 
Throughout trapped inflation in higher dimensions, $\chi$-particles with a $\vec{\varphi}$-dependent mass are present. The larger the mass of a particle, the further away from the current position on the trajectory is the ESP at which the particle has been produced. As this distance in field space grows, particles at a given ESP become more and more diluted due to inflation while their mass increases. As particles become more massive, their decay rate to lighter degrees of freedom increases, $\Gamma_\chi \propto m$ (i.e. $\Gamma_{\chi}=h^2m/(8\pi)$ in the presence of a coupling to a spinor field $-h\psi\bar{\psi}\chi$ \cite{Kofman:1997yn}), and they decay once $\Gamma_\chi$ becomes of order $H$. If they decay while their density is still large enough so that they can be described by a homogeneous field (as we assumed throught the paper), they will leave no additional observational traces, since $\rho_\chi\ll v_t^2/2\ll \rho_\varphi$. Note that they need to be stable for at least a few Hubble times after their creation in order for backreaction to become important in the first place. 

On the other hand, if $\chi$-particles are stable enough that their number density becomes exceedingly low before they decay, we have to consider the effect of individual, massive particles (a few per Hubble-horizon) that are directly coupled to the inflatons. In the vicinity (in real space) of such a particle the inflatons are held back. As a result, the energy-density $\rho_\varphi$ in a spherical region around the particle is slightly increased. This in turn delays reheating and causes a circular cold spot in the CMBR with an amplitude that is proportional to the induced over-density \footnote{We would like to thank A.~Dahlen for pointing out this possibility.}, see i.e.~\cite{Itzhaki:2008ih,Fialkov:2009xm,Kovetz:2010kv}. Since these massive $\chi$-particles are resupplied throughout trapped inflation, we arrive at an additional source of perturbations, consisting of the superpositions of spherical over-densities. We expect these additional signatures to be ill suited to provide the dominant contribution to the power-spectrum. An investigation of this interesting effect is in preparation.

\subsection{A Bending Field Trajectory \label{sec:meander}}

The force due to backreaction of $\chi$-particles is not anti-parallel to the velocity if the trajectory is curved on length-scales of order $v_t/(3H)$, where $1/(3H)$ is the time-scale over which backreaction of particles at a given ESP is important. As a result, any bending of the trajectory is enhanced,  causing the trajectory to twist and turn. Inflation may still take place, but since the trajectory bends considerably during any given Hubble time, isocurvature fluctuations constantly feed into the adiabatic mode, making the power-spectrum less scale invariant and enhancing non-Gaussianities. In order to say anything more concrete, we need to specify the classical potential that determines the classical trajectory. 

\subsubsection{A Meandering Path} 
Assume that the inflaton potential is smooth on large scales with a constant average gradient in the direction of $\varphi_1$, but with random small scale bumps and dips on top. Whenever the trajectory encounters a bump it gets disturbed, causing it to find a way around the bump. The resulting trajectory is reminiscent of the path a meandering river takes flowing down a mountain side, see i.e \cite{Li:2009sp,Li:2009me,Wang:2010rs} for related proposals. 

On such a potential, the trajectory encounters many bifurcation junctions: for example, if $D=2$, the trajectory could go to the left or right of a bump. Which way is chosen depends sensitively on the initial conditions. If we deal with many bumps, the displacement perpendicular to $\varphi_1$ can be described by a random walk, whereas $\varphi_1$ remains dominated by the averaged gradient. If $D\geq 4$ the probability that trajectories reunite approaches zero, since a random walk is not self intersecting in three or more dimensions.

Inflation in such a model is troublesome: at best, the universe is covered by domain walls after inflation if by some miracle in all the different inflationary Hubble patches that are observable today inflation terminated and reheating commenced simultaneously to within a fraction of a Hubble time, even though the trajectories in each patch are different. Without such a miracle, inflation ends at different times, leading to pronounced hot/cold spots in the CMB, that are not observed. 

Adding ESLs and backreaction to such a model enhances these problems. Therefore, we do not consider meandering trajectories further.

\subsubsection{A Curved Path}
Due to the problems caused by bifurcation junctions, such as domain walls and pronounced hot/cold spots in the CMBR, we would like to shift attention to potentials that channel the trajectory along a single path throughout the last sixty e-folds of inflation. This means that we would like to investigate potentials that have a positive semidefinite slope in all directions perpendicular to the trajectory, see for example \cite{Berg:2009tg} for a recent proposal of an in-spiraling trajectory in two dimensions.

If we consider models with a clear channel, that is with slopes within the channel walls that are bigger than the slope along the channel, and if the classical trajectory is only curved but not intertwined over length-scales of interest, we expect the effect of backreaction to be still most pronounced in the direction opposite to the velocity. Thus, we still expect the presence of a terminal velocity, as derived in Sec.~\ref{sec:particpleproductionwithimpactparameter}. If backreaction alters the trajectory drastically, i.e. by forcing the fields over a channel-wall, viable inflationary models could still result, but little can be said generically. We leave a numerical investigation of concrete models to future studies.

\section{Conclusion}
Whether or not moduli get trapped at extra species loci (ESL), which are often associated with additional symmetries, has important consequences for selecting vacua on the landscape. If the trajectory in field space comes close to such a locus, particles are produced and backreaction can trap the trajectory. In this paper we gave a simple geometric argument to highlight that this type of moduli trapping is unlikely if the dimensionality of moduli space is considerably larger than the dimensionality of the loci; for $D>d+1$ (or $D>d+2$ if the trajectory is intertwined), the probability of getting trapped is suppressed by $(\mu/x)^{D-d-1}$ (or $(\mu/x)^{D-d-2}$ if the trajectory is intertwined), where $x$ the characteristic inter-ESL distance, $\mu=\sqrt{v/g}<x$ is an effective trapping radius, $v$ the velocity in field space, and $g$ a coupling constant. We concluded that the sole presence of ESLs does not guarantee a dynamical preference of high symmetry vacua on the landscape, but since the effective trapping radius increases with the velocity, ESLs are always crucial at large speeds, for instance if moduli are responsible for inflation.

For $D\gg 1$ and $d=0$ (extra species points or ESPs, a generalization to ESLs is straightforward), we derived a terminal velocity $v_t=g^2x$ on the landscape: once the speed approaches $v_t$, the trajectory comes within reach of a plethora of ESPs at which $\chi$-particles are produced. Their combined backreaction acts opposite to the velocity if the classical trajectory is straight over distances covered in a few Hubble times (we allow for a classical potential of the moduli). Since backreaction increases sharply for $v>v_t$ if the dimensionality of moduli space is large, $v_t$ acts as a speed limit. The classical trajectory is followed with this speed as long as the slope along the classical potential is large and the path is reasonably straight. 

The presence of a terminal velocity offers the intriguing possibility to drive inflation with potentials that do not otherwise support slow-roll inflation. This type of inflation is a generalization of trapped inflation. The main difference is the independence of the speed from the slope of the potential, as long as $D$ is large enough. 

We then provided a phenomenological model of trapped inflation in higher dimensions, assuming a simple quadratic potential and ubiquitous ESPs ($x\lesssim 10^{-3.5}$, $g\sim 0.1$). For these parameters, more than sixty e-folds for inflation result for sub-Planckian excursions in field space and observations are within observational bounds (the power-spectrum of curvature perturbations is slightly red and gravitational waves are suppressed). We would like to highlight that no fine tuning of initial conditions is needed and constraints on the potential are lessened (the COBE normalization still sets the inflationary scale, but the $\eta$-problem is alleviated). Additional observational signatures are expected, such as an extra contribution to the power-spectrum from IR-cascading (less than a few percent if $g<0.2$), potentially large non-Gaussianities and additional circular cold-spots in the CMBR caused by individual $\chi$-particles that became heavy before being diluted away. We plan to come back to these interesting signatures in a future study. 

In this paper, we took a bottom-up approach and parametrized the undoubtedly rich structure of the landscape and the distribution of ESLs by a few characteristic parameters, such as $x$, that we left undetermined. Concrete implementations within string theory are needed to see whether or not the distribution of ESLs and the potential on the landscape supports trapped inflation in higher dimensions. It appears that ubiquitous ESLs and steep potentials are common, but we leave an in depth analysis to future studies.

 %%%%%%%%%%%%%%%%%%%%%%%%%%%%%%%%%%%%
\begin{acknowledgments}
We are especially grateful to S.~Watson for extensive discussions, encouragement and comments on the draft. We would also like to thank N.~Barnaby, A.~Brown, A.~Dahlen, L.~Kofman, J.L.~Lehners, L.~Lorenz, S.~Patil and P.~Steinhardt for discussions. T.B.~is supported by the Council on Science and Technology at Princeton University and is grateful for the hospitality of Michigan University, the Perimeter Institute and CITA.

\end{acknowledgments}
%%%%%%%%%%%%%%%%%%%%%%%%%%%%%%%%%%%%

\appendix

\section{Trapping Probability, Improved Estimate  \label{sec:volumeargument}}
It is instructive to estimate the trapping probability by means of a volume argument, which does not require any artificial shifting of ESLs.  Consider the volume in the vicinity of the classical trajectory such that the shortest distance to $\vec{\varphi}(t)$ is less than $\mu$
\begin{eqnarray}
V_{cl}\equiv sV_{D-1}\mu^{D-1} \,.
\end{eqnarray}
If we consider ESPs (d=0), we can estimate the probability of trapping simply as $p_{cl}\approx V_{cl}/x^D=V_{D-1}(s/x)(\mu/x)^{D-1}$. Note that the numerical pre-factor differs by  $1/D$ if we compare the probability with (\ref{probtarpesp}); the origin of  $1/D$ in (\ref{probtarpesp}) can be traced to shifting  ESPs to the surface, so its absence  is expected. 

We can generalize this argument again to higher dimensional ESLs, but we need to specify the orientation of ESLs in order to derive analytic expressions for the trapping probability \footnote{We did not need to specify the orientation in section \ref{sec:classicalESL}, since we treated ESLs as if they were laid out over the surface of the hyper-sphere with radius $s$ around $\vec{\varphi}(0)$; consequently, section \ref{sec:classicalESL} is more general with respect to the orientation of the ESLs, but our arguments depend on the artificial movement of ESLs to the surface, which leads to a lower bound of the probability.}: consider a periodic grid of ESLs ($d\geq 1$) with a grid-spacing of $x$, such that one extension of an ESL is in the same direction as $\vec{\varphi}(t)$. For example, if $d=1$ and $D=3$, we consider a cubic lattice where any given extra species lines intersect with two other extra species lines at right angles after a distance $x$, and one class of lines is parallel to $\varphi (t)$. Such an arrangement is  artificial, but it allows for a simple estimate of $p_{cl}$ that provides an approximation for more random distributions of ESLs that share the same inter-ESL distance $x$.

\begin{figure*}[tb]
\includegraphics[scale=0.4,angle=0]{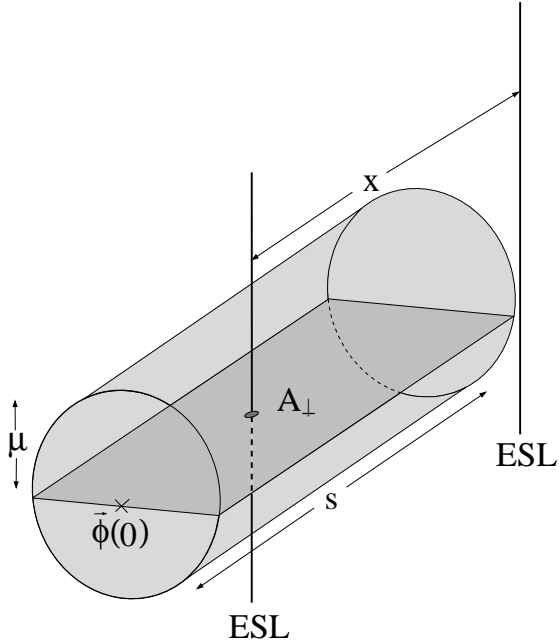}
   \caption{\label{pic:test} Schematic in $D=3$, showing $V_{cl}=sV_{D-1}\mu^{D-1}$ (the gray cylinder), two (exemplary) extra species lines ($d=1$) separated a distance $x$ that are perpendicular to the classical trajectory $\vec{\varphi}(t)$ (not shown) of length $s$; $\varphi (t)$ lies in the middle of the dark gray rectangle (area $A_{\perp}=s\mu^{D-1-d}V_{D-1-d}=2\mu s$), connecting the two faces of the cylinder. The probability of getting trapped is $p_{cl}\approx N_{\perp}A_{\perp}/x^{D-d}=4\mu s/x^2$, in accordance with (\ref{trappingprobabilityclassical}).}
\end{figure*}

Given such a grid, there are
\begin{eqnarray}
N_{||}=\left( {{D-1}\atop {d-1}}\right)=\frac{(D-1)!}{(d-1)!(D-d)!}
\end{eqnarray}
possible orientations of an ESL with one direction parallel to $\vec{\varphi(t)}$ and 
\begin{eqnarray}
N_{\perp}=\left( {{D-1}\atop {d}}\right)=\frac{(D-1)!}{d!(D-1-d)!}\label{Nperp}
\end{eqnarray}  
possible orientations perpendicular to $\vec{\varphi}(t)$.
The probability for a parallel (perpendicular) orientation to intersect $V_{cl}$ is
\begin{eqnarray}
p_{||}&=&N_{||}V_{D-d}\mu^{D-d}\frac{1}{x^{D-d}}\,,\\
p_{\perp}&=&N_{\perp}sV_{D-1-d}\mu^{D-1-d}\frac{1}{x^{D-d}}\,.
\end{eqnarray}
For example, if $d=1$ and $D=3$ (so that $N_{||}=1$ and $N_{\perp}=2$), we can identify $1/x^2$ as the number density of intersection points between the extra species lines and projections of $V_{cl}$ onto planes parallel or perpendicular to $\vec{\varphi}(t)$; to be concrete, $sV_{D-1-d}\mu^{D-1-d}=V_{1}\mu s=2\mu s$ is the cross section of $V_{cl}$ along $\vec{\varphi}(t)$; thus, $2\mu s/x^2$ counts the number of intersection points of the ESLs with the cross-section. This is identical with the probability of being trapped at one of these ESLs if $2\mu s/x^2<1$. See Fig.~\ref{pic:test} for a schematic.

The total trapping probability is given by the sum $p_{cl}= p_{\perp}+p_{||}$, but since $\mu \ll s$ we can approximate 
\begin{eqnarray}
p_{cl}\approx p_{\perp}=N_{\perp}V_{D-1-d}\frac{s}{x}\left(\frac{\mu}{x}\right)^{D-1-d}\,.\label{trappingprobabilityclassicalapp}
\end{eqnarray}
Note that this expression is only applicable for $p_{cl}<1$.
If we compare this probability with the lower bound in (\ref{trappingprobabilityclassicallb}), we observe again the absence of the $1/D$ suppression (as expected), and a numerical pre-factor $N_{\perp}$, which is determined by the chosen orientation of ESLs along a grid. For comparison, if all ESLs were parallel to each other, while having no extent in the direction of $\vec{\varphi}(t)$, the pre-factor would be $N_{\perp}=1$. Ignoring the pre-factor, which needs to be determined from the given distribution of ESLs in concrete cases, we find the general scaling
\begin{eqnarray}
p_{cl}\propto V_{D-1-d}\frac{s}{x}\left(\frac{\mu}{x}\right)^{D-1-d}\,.
\end{eqnarray} 
For large $D-1-d$, we observe a strong suppression due to  $(\mu/x)^{D-1-d}$ as well as $V_{D-1-d}$. We conclude that moduli trapping by ESLs is only efficient if $d=D-1$, that is  ESPs can trap moduli efficiently on a line ($D=1$), but not on a plane ($D=2$), or higher dimensional spaces.

\section{Examples \label{sec:examples}}
\subsection{Size Moduli}
It has been argued \cite{Watson:2004aq,Patil:2005fi} (see \cite{Battefeld:2005av} for a review) that the size of internal dimensions can be stabilized by string gases that contain enhanced symmetry states \footnote{See \cite{Cremonini:2006sx} for a related work in M-theory.}. String Gas Cosmology (SGC) is a low energy field theoretical approach incorporating strings and branes in a cosmological setting by means of a gas approximation. If internal dimension come close to the self dual radius, certain combinations of winding and momentum modes become massless. Once they are produced in sufficient quantities, their backreaction can stabilize internal dimensions temporarily, up until the expansion of the universe dilutes them away. 

There are six internal dimensions in string theory, $D=6$, and the light states appear whenever any of the six internal dimensions comes close to the self dual radius, $d=D-1=5$; because $d=D-1$, this type of trapping is a simple example of an unsuppressed trapping probability. As a consequence, temporary trapping appears unavoidable if an internal dimension approaches the self dual radius. 

\subsection{Preheating \label{sec:examplepreheating}}
Preheating refers to a possible phase of explosive particle production after inflation, whereby a large fraction of the energy in the inflaton fields is transferred to other degrees of freedom. There are several phenomenological models of preheating employing resonances \cite{Traschen:1990sw,Kofman:1997yn} or instabilities \cite{Felder:2000hj,Dufaux:2006ee}, see \cite{Bassett:2005xm,Kofman:2008zz} for reviews and comprehensive referencing.

Simple parametric resonance models operate via the mechanism described in Sec.~\ref{sec:trapping}, with $\varphi_i$ ($i=1\dots D$) identified as the inflatons. These are coupled to a ``matter''-field $\chi$, which is assumed to decay further to ordinary standard model particles. Let's assume that all inflaton fields couple to one matter field only, which becomes light at $\varphi_i=0$. Because the condition for resonant particle production is (\ref{defmu})
\begin{eqnarray}
\sum_i\varphi_i^2 \lesssim \frac{v}{g}\,, 
\end{eqnarray}
we conclude that all inflatons need to get close to the origin simultaneously (in phase), in order for parametric resonance to take place. Even if the inflaton potential has a global minimum at $\vec{\varphi}=0$, it is unlikely that all fields start out in phase and oscillate with the same frequency. Because of the generic dephasing of fields, we expect parametric resonance to be negligible if several inflatons couple to the same matter field. This is indeed the case, as shown in \cite{Battefeld:2008bu,Battefeld:2008rd} analytically at the example of $\mathcal{N}$-flation, and by means of lattice simulations in \cite{Battefeld:2009xw}, where it was also shown that for as little as $D=3$, parametric resonance is heavily suppressed (see also \cite{Barnaby:2010wd}). This is in agreement with our conclusion that trapping events are unlikely if $d<D-1$, see eqn.(\ref{trappingprobabilityclassicallb}). One can still observe resonances for $D=2$ \cite{Battefeld:2009xw} even though $0=d<D-2=1$, since the classical potential of the inflatons, which we neglected in deriving (\ref{trappingprobabilityclassicallb}), leads to repeated attempts of particle production at the same ESP \footnote{The case $D=2$ is even more subtle: in \cite{Bassett:1997gb,Bassett:1998yd} it was shown that almost all Fourier modes of $\chi$ get amplified if the oscillating frequencies of the inflatons are not simple ratios. This enhancement is sometime referred to as Cantor preheating and was shown to be inefficient for $D\geq 3$ in \cite{Battefeld:2009xw}, see also \cite{Barnaby:2010wd}.}. Note that the absence of parametric resonance may be desirable: since the reheating temperature is lowered considerably, the overproduction of thermal relics, such as gravitinos \cite{Ellis:1984eq,Khlopov:1984pf,Kawasaki:2008qe}, is avoided.

In \cite{Battefeld:2009xw} it was also shown that tachyonic instabilities, that is explosive particle production caused by Yukawa interactions of the inflatons to the matter field ($\mathcal{L}_{int}=\sum_i \sigma_i\varphi_i\chi^2/2$), are left intact if the number of fields is increased. Such interactions are not present in the cases we consider in this paper.

\section{Drifting through Moduli Space during Inflation \label{A:trappingrw}}
During inflation any classical movement of moduli is quickly damped by Hubble friction, as we have seen in section \ref{sec:length}. However, quantum mechanical fluctuations cause the fields to drift in random directions \footnote{Here and in the following we assume that Hubble induced corrections to the moduli's mass are negligible. If they are present, the fields do not drift further than $\delta \varphi_i\sim H$.}. Consider a single modulus located initially at the origin $\varphi(0)=0$, with negligible classical potential (the moduli we consider here are not the degrees of freedom that drive inflation; see section \ref{sec:moduliareinflatons} for the latter case). Inside the horizon, quantum fluctuations displace the field from $\varphi=0$ all the time; these fluctuations become classical shifts of the field after horizon crossing. Thus, if we look within our Hubble patch and ask what is the value of $\varphi$, homogeneous within our Hubble patch, we generically find a non-zero value. If we were to look at an ensemble of Hubble patches, which inflation creates, we expect to see a Gaussian distribution. Following \cite{Linde:2005ht}, the probability distribution of finding a coordinate volume filled with $\varphi \neq 0$ satisfies a diffusion equation \cite{Linde:2005ht} 
\begin{eqnarray}
\frac{\partial P(\varphi,t)}{\partial t}=\mathcal{D}\frac{\partial^2 P(\varphi,t)}{\partial \varphi^2}\,,
\end{eqnarray}
where $\mathcal{D}=H^3/8\pi^2$, which can easily be solved for $H\sim const$, yielding the expected Gaussian probability distribution
\begin{eqnarray}
P(\varphi,t)=\frac{2\pi}{H^3t}\exp\left(-\frac{2\pi^2\varphi^2}{H^3t}\right)\,.
\end{eqnarray}  
Thus, the expectation value of $\varphi_i^2$ in a Hubble volume is 
\begin{eqnarray}
\left< \varphi^2 \right> &=&\int\varphi^2 P(\varphi,t)\,d\varphi\\
&=&\frac{H^3}{4\pi^2}t\,.
\end{eqnarray} 
Since the number of e-folds during inflation is 
\begin{eqnarray}
N=\int H\, dt\approx H t\,
\end{eqnarray}
we get the rms value of the modulus after $N$ e-folds
\begin{eqnarray}
\varphi^{rms}\equiv \sqrt{\left< \varphi^2 \right>}=\sqrt{N}\frac{H}{2\pi}\,.
\end{eqnarray}
If we are not interested in resolving  the field trajectory on smaller scales than $H/(2\pi)$, we can approximate the diffusive, continuous drifting of the field in any given Hubble patch by a random walk, whereby the field makes a step to the left or right of length $H/(2\pi)$ after each e-fold of inflation. 

In a higher dimensional moduli-space, each of the fields diffuses independently of the other ones, causing a random walk in the higher dimensional space; after each e-fold,  a step of length $l=H/{2\pi}$ is performed in a random direction. To keep things simple, we approximate the continuous choice of direction by a discrete one along a D-dimensional cubic lattice along the $\varphi_i$-directions. This approximation is viable if we do not resolve the path on scales comparable or smaller than the step size.

Given such a random walk in field space \footnote{We work with a non-compact field space in this section. If moduli space is compact and the trajectory is ergodic \cite{Horne:1994mi}, one may still use the arguments in this section by treating the field space as a covering space of the moduli space.}, 
there are several effects determining the trapping probability:
\begin{itemize}
\item The maximal length of the random walk in $D$-dimensions, which is given by the largest radius of gyration $R_1$ (see section \ref{sec:shapeofarandomwalk}). Using $s=s_{rw}=2R_1$ in (\ref{trappingprobabilityclassical}) yields a lower bound on the trapping probability.
\item The shape and volume coverage of a random walk. Even though the ensemble average over many random walks has spherical symmetry in moduli space, an individual random walk is far from spherical. In the large $D$ limit, one can derive analytic expressions for all radii of gyration (see section \ref{sec:shapeofarandomwalk}). Within the volume spanned by the radii of gyration, the random walk has an increased probability of getting close to an ESL (the Hausdorff dimension of a random walk for $D\geq 2$ is two; thus, it is more volume filling than a straight trajectory with the same $s$).   
\item  Expansion due to inflation. In an inflating universe, one might consider not only the history of our own Hubble volume, but the ensemble of all Hubble patches that became causally disconnected during inflation. Since the moduli perform an independent random walk in each causally disconnected Hubble patch, one might be tempted to include an additional volume factor that enhances the probability. We argue in section \ref{sec:expansioninflation} that this line of reasoning is flawed.
\end{itemize}
In the following, we assume $\sqrt{N}l>x>\mu$. 

\subsection{Random Walks in D Dimensions \label{sec:shapeofarandomwalk}}
Consider a random walk of step-length $l=H/(2\pi)$ on a lattice in a D-dimensional space. Given a walk consisting of $N$ steps ($N$ is identical to the number of e-folds in our case) we can write the position after the k'th step as $\vec{\varphi}(k)$, that is the i'th modulus field has a magnitude of  $\varphi_i(k)$ after the k'th step. A  walk's shape  can be characterized  by the radius of gyration tensor, defined as \cite{Rudnick:2004,Rudnick:87}
\begin{eqnarray}
R_{ij}\equiv \frac{1}{N}\sum_{k=1}^{N}\left(\varphi_i(k)-\left< \varphi_i \right>_{rw}\right)\left(\varphi_j(k)-\left<\varphi_j\right>_{rw}\right) \label{rtensor}\,,
\end{eqnarray} 
in analogy to the moment of inertia tensor in classical mechanics. Here, $\left<.\right>_{rw}$ is the average over a particular random walk, that is
\begin{eqnarray}
\left<\varphi_i\right>_{rw}\equiv\frac{1}{N}\sum_{k=1}^{N}\varphi_i(k) \,.
\end{eqnarray}
The eigenvalues of (\ref{rtensor}), $R_j^2$, provide an estimate of the walk's extent in the direction of the corresponding eigenvector. One can derive analytic expressions for the ensemble average $\left<.\right>_{en}$ in a $1/D$-expansion (which becomes exact in the limit of $D\rightarrow \infty$), leading to the radii of gyration  
\begin{eqnarray}
\left<R_j^2\right>_{en}\approx l^2 \frac{N+1}{\pi^2 j^2}\left(1+\frac{3}{4D}\right)\,, \label{radiiog}
\end{eqnarray} 
up to first order in $1/D$ ($j=1\dots D$).
The associated variance is
\begin{eqnarray}
\left<\left(R_j^2-\left<R_j^2\right>_{en}\right)^2\right>_{en}\approx \frac{2l^2}{D}\left(\frac{N+1}{j^2\pi^2}\right)^2\,,
\end{eqnarray} 
which becomes smaller with increasing $D$. It should be noted that the error of the $1/D$ expansion is smallest for low $j$, and radii are better approximated than variances \cite{Rudnick:2004}. However, even for $D=3$ and keeping  terms up to order $1/D$, the approximation is accurate enough for our purposes; mean values of $R_1^2$ from simulations agree with the ensemble average of the eigenvalue to within $1\%$, $R_2^2$ to within $10\%$ and $R_3^2$ to within $25\%$. These errors are consistent with expected corrections of order $1/D^2$.  

As evident from the radii of gyration in (\ref{radiiog}), the shape of an individual random walk is far from spherical, since a sphere has $R_j=R$ for all $j$. One can quantify the amount of asphericity by
\begin{eqnarray}
A_D&\equiv & \frac{\sum_{i>j}^{D}\left<\left(R_i^2-R_j^2\right)^2\right>_{en}}{(D-1)\left<\left(\sum_{i=1}^{D}R_i^2\right)^{2}\right>_{en}}\\
&=&\frac{4+2D}{4+5D}\,,
\end{eqnarray}
a number that is between zero (a sphere) and one (a stick). For large $D$, the asphericity approaches $2/5$, indicating that random walks in higher dimensions are neither stick-like, as the classical trajectory, nor spherical, as an average over many random walks. Further, they  all ``look'' alike, regardless of the dimensionality D. 

\begin{figure*}[tb]
\includegraphics[scale=0.4,angle=0]{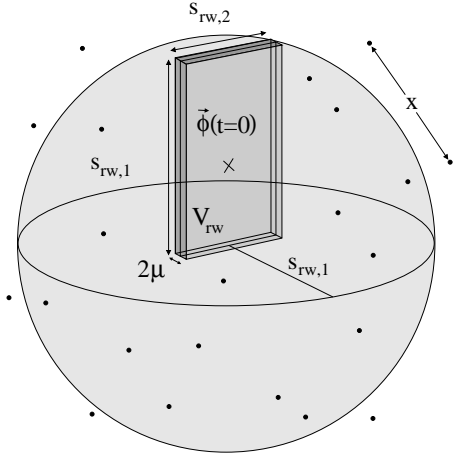}
   \caption{\label{pic:sphere} Visualization of the volume sampled by a random walk in $D=3$: the random walk covers an area that can be estimated by the two largest radii of gyration $A_{rw}=s_{rw,1}s_{rw,2}\sim 4 R_1R_2$, where $s_{rw,i}$ are given by twice the ensemble average of the corresponding radius of gyration. Moduli get trapped if an ESL intersects the volume $V_{rw}=A_{rw}V_{D-2}\mu^{D-2}=2A_{rw}\mu$ (the grey monolith in the picture). Also shown are ESPs (d=0) with a characteristic separation of $x$, so that $p_{rw}\approx V_{rw}/x^D=2s_{rw,1}s_{rw,2}\mu/x^3$ (only applicable for $p_{rw}<1$).}
\end{figure*}

The random walk is spread over
\begin{eqnarray}
s_{rw,j}&\equiv & 2\sqrt{\left< R_j^2\right>_{en}}\\
&\approx& 2l\frac{\sqrt{N+1}}{\pi j}\left(1+\frac{3}{4D}\right)^{1/2}\,.
\end{eqnarray} 
Hence, the largest extent of the random walk is given by
\begin{eqnarray}
s_{rw}&\equiv& s_{rw,1}\approx 2l\frac{\sqrt{N+1}}{\pi}\left(1+\frac{3}{4D}\right)^{1/2}\,. \label{srw}
\end{eqnarray}

A naive guess for the volume sampled by a random walk might be $\Pi_{j=1}^{D}(s_{rw,j})$. However, this does not properly take into account the volume coverage of a random walk; its fractal dimension, that is its Hausdorff dimension, is 2 for all $D\geq 2$. Thus, a random walk is not volume filling, but it covers an analog of a two dimensional surface with an area of order $s_{rw,1}s_{rw,2}$. Since we are interested in cases where the random walk comes to within a distance of order $\mu$ to an ESL, we may define an effective volume of (see Fig.~\ref{pic:sphere})
\begin{eqnarray}
V_{rw}&\equiv & V_{D-2}\mu^{D-2} s_{rw,1}s_{rw,2}\\
&=&l^2 \mu^{D-2} V_{D-2}6\left(1+\frac{3}{4D}\right)\frac{(N+1)}{\pi^2}\,. \label{volumerw}
\end{eqnarray}
$V_{rw}$  can be used in estimating the trapping probability, see section \ref{sec:trappingrandomwalk}.

\subsection{Trapping a Random Walk}
Moduli trapping, and the magnitude of $\mu$, differ from the process described in section \ref{sec:trapping}, since the movement of the random walk is not driven by kinetic energy, but by quantum fluctuations. The description in section \ref{sec:trapping} is based on modeling the trapping event in field theory by the inclusion of a classical field $\chi$, which acquires a non-zero VEV due to explosive particle production once the trajectory in moduli space comes closer than $\mu$ to an ESL. In this process, part of the kinetic energy in the moduli is transferred to $\chi$, which then backreacts onto the $\varphi_i$. However, if the moduli simply drift through moduli space, driven by quantum mechanical fluctuations, this process can not be operational. 

Nevertheless, there is still a mechanism that can trap moduli: let's denote the distance from an ESP after which a field theoretical description with an additional scalar field $\chi$ becomes viable by $\mu_{ft}$. Once the trajectory comes closer than $\mu_{ft}$ to the ESP, we also have to consider the effect of quantum mechanical fluctuations on $\chi$, which increases $\chi_{rms}\equiv \sqrt{<\chi^2>}$. In addition to this quantum mechanical drift, the field is affected by the classical potential 
\begin{eqnarray}
V=\frac{g^2}{2}\chi^2\sum_{i=1}^{D}\varphi_i^2\,.
\end{eqnarray}
Since $\sum_i\varphi_i^2\approx \mu_{ft}^2$, a classical force driving $\chi$ back to the origin results. To be concrete, if the usual slow-roll approximation for $\chi$ is used, which is justified since we start out at $\chi=0$, one finds \cite{Linde:2005ht,Allahverdi:2007wh} a modified diffusion equation for the probability distribution of finding $\chi$ at a nonzero VEV,
\begin{eqnarray}
\frac{\partial P_{\chi}}{\partial t}=\frac{H^3}{8\pi^2}\frac{\partial^2 P_{\chi}}{\partial \chi^2}+\frac{1}{3H}\frac{\partial}{\partial \chi}\left(P_{\chi}\frac{\partial V}{\partial \chi}\right)\,,
\end{eqnarray}
where the last term originates from the classical force on $\chi$. This leads to an equation of motion for $\left< \chi^2 \right>$
\begin{eqnarray}
\frac{d\left<\chi^2 \right>}{dt}=\frac{H^3}{4\pi^2}-\frac{2}{3H}\left<\frac{\partial V}{\partial \chi}\chi \right>\,.
\end{eqnarray}
Since $\partial V/\partial \chi \approx g^2 \mu_{ft}^2\chi$, we can integrate the above to
\begin{eqnarray}
\left<\chi^2\right>=\frac{3H^4}{8\pi^2 g^2\mu_{ft}^2}\left(1-\exp\left(-\frac{2g^2\mu_{ft}^2}{3H}t\right)\right)\,.
\end{eqnarray}
After $N\gtrsim 3H^2/(2\mu_{ft}^2)$ e-folds of inflation, the value of $\chi_{rms}$ settles into
\begin{eqnarray}
\chi_{rms}=\sqrt{\frac{3}{8\pi^2}}\frac{H^2}{g^2\mu_{ft}^2}\,.
\end{eqnarray}
This in turn introduces a classical force onto the moduli $\propto \partial V/ \partial \varphi_i\sim g^2\chi_{rms}^2\varphi_i$, driving them towards the position of the ESP. Once they get closer to the ESP, $\chi$ can drift out further, increasing the classical force on the moduli, etc. In the end, the moduli drift around the ESP, while $\chi$ fluctuates around zero, and none of them can escape. Thus, the moduli are trapped near the ESP (a generalization to higher dimensional ESLs is straightforward). 

Interestingly, the effective trapping radius is given by $\mu=\mu_{ft}$, which is necessarily bigger than the value in section \ref{sec:trapping}, since the arguments in section \ref{sec:trapping} are also based on a field theoretical description. Thus, trapping by ESLs is more efficient for drifting moduli than for classically moving ones. If $\mu_{ft}$ is bigger than the expected inter-ESL distance $x$, we conclude that moduli are preferably located near ESLs after inflation, given that inflation lasts sufficiently long so that $s_{rw}\gg \mu_{ft}$.

The value of $\mu_{ft}$ is model dependent; for example, if the moduli correspond to distances between D-branes and the light extra degrees of freedom correspond to strings stretching between them, then the strings are only light, and can thus be modelled in field theory by $\chi$, if the branes come to within the fundamental string length. This in turn determines $\mu_{ft}$. 

In the following, we leave $\mu=\mu_{ft}$ general and take $s_{rw}>x>\mu$ to estimate the trapping probability if $\mu_{ft}$ turns out to be less than $x$.

\subsection{Trapping Probability \label{sec:trappingrandomwalk}}
Using the maximal extent of the random walk $s_{rw}$ from (\ref{srw}) in (\ref{trappingprobabilityclassical}) yields a lower bound on the trapping probability

\begin{eqnarray}
p_{rw}>p_{cl}(s_{rw})&\approx& N_{\perp}V_{D-1-d}\frac{s_{rw}}{x}\left(\frac{\mu}{x}\right)^{D-1-d}\\
&=&N_{\perp}V_{D-1-d}\frac{2}{\pi}\left(1+\frac{3}{4D}\right)^{1/2}\frac{\sqrt{N+1}l}{x}\left(\frac{\mu}{x}\right)^{D-1-d} \label{lowerboundprw}
\end{eqnarray}
where $l=H/(2\pi)$ and $N=\int H dt\approx Ht$ is the number of e-folds, identical with the number of steps in the random walk. 

Incorporating the increased volume coverage of the random walk leads to a better estimate of the trapping probability. Analogously to appendix \ref{sec:volumeargument}, we distribute the ESLs along a grid. The total probability is now a sum of four parts, $p_{\perp\perp}$ (all ESL directions are perpendicular to $s_{rw,1}$ and $s_{rw,2}$), $p_{\perp\,||}$ and $p_{||\,\perp}$ (one ESL direction is parallel to $s_{rw,1}$ or $s_{rw,2}$, while all other ones are perpendicular) and $p_{||\,||}$ (one direction is parallel to $s_{rw,1}$ and an other one to $s_{rw,2}$). Each probability is given by the appropriate cross-section of $V_{rw}$ (such that the ESLs appear as points in this cross-section) multiplied with $1/x^{D-d}$ and the number of possible orientations of an ESL within this class. Since $\mu\ll s_{rw,1},s_{rw,2}$, we can approximate the total probability by $p_{rw}\approx p_{\perp \,\perp}$. If ESLs are distributed along a grid, there are
\begin{eqnarray}
N_{\perp\,\perp}=\left( {{D-2}\atop {d}}\right)
\end{eqnarray} 
possible orientations perpendicular to both $s_{rw,1}$ and $s_{rw,2}$, and the probability becomes
\begin{eqnarray}
p_{rw}&\approx& N_{\perp\,\perp}V_{D-2-d}\mu^{D-2-d}s_{rw,1}s_{rw,2}\frac{1}{x^{D-d}}\\
&=&N_{\perp\,\perp}V_{D-2-d}\frac{2}{\pi^2}\left(1+\frac{3}{4D}\right)\frac{(N+1)l^2}{x^2}\left(\frac{\mu}{x}\right)^{D-2-d}\,. \label{probabilityrandomwalk}
\end{eqnarray}
As before, a different distribution of ESLs would change the pre-factor, but not the general scaling with $\mu/x$.
A comparison with (\ref{lowerboundprw}) reveals an enhancement of $\sqrt{N+1}l/\mu$, caused by the higher volume coverage of the random walk. However, for large $D-2-d$, we still observe a strong suppression. We conclude, that moduli trapping at ESLs in $D$ dimensions is expected to be efficient during inflation only if $d\geq D-2$; for instance,  lines can trap efficiently in three dimensions, while points can't. 

The expression for $p_{rw}$ in (\ref{probabilityrandomwalk}) is also valid for a classical, intertwined trajectory, that can be approximated by a random walk. In this case $\mu=\sqrt{v/g}$, $N$ is still the number of steps (but not the number of e-folds) and $l$ the step length, all of which  need to be determined from the classical trajectory.

\subsection{Expansion Effects \label{sec:expansioninflation}}
If a random walk in moduli space begins in some Hubble patch during inflation, $e^3$ identical copies of this Hubble patch appear after the first step (that is after one e-fold), and the moduli perform independent random walks in each patch from then on.

Consider now a random walk of length $s_{rw,1}=x$. The volume  sampled by a single walk of this length is (\ref{volumerw})
\begin{eqnarray}
V_{rw}=\frac{x^2}{2}\mu^{D-2}V_{D-2}\,.
\end{eqnarray}
Thus, it would take about
\begin{eqnarray}
n\equiv \frac{x^DV_D}{V_{rw}}=\frac{2x^{D-2}V_D}{\mu^{D-2}V_{D-2}}
\end{eqnarray}
independent random walks to cover the entire volume of the D-dimensional sphere of radius $x$. 
As a consequence, the probability of finding at least one Hubble patch where the moduli get trapped at an ESL approaches one, if inflation lasted $N_{add}>\ln(n)/3$ e-folds in addition to the 
number of e-folds needed to achieve $s_{rw}=x$.

Thus, one might be tempted to conclude that inflation enhances the chance of being trapped by ESLs. However, this line of thought is flawed: the volume enhancement factor does not prefer ESLs; inflation allows for the population of all vacua that are within reach of a random walk, not only those close to an ESL. Hence, the presence of ESLs is irrelevant if arguments to select a vacuum are contingent on this indiscriminatory population of moduli space. 

An additional problem with choosing vacua randomly as above is the presence of domain walls at late times: if the fields drift through moduli space during the last sixty e-folds of inflation, different regions of the universe today have distinct values for the moduli-fields. In this section, we ignored any classical potential for the $\varphi_i$, an approximation that may be justified during high scale inflation but certainly not today. As a consequence, our universe would be filled with domain walls separating regions with different values for the $\varphi_i$, a setup that is ruled out observationally.

\end{document}